\documentclass[twocolumn,english,aps,prl,twocolumn,showpacs,superscriptaddress,groupaddress,floatfix,graphics,graphicx]{revtex4-1}
\usepackage[T1]{fontenc}
\usepackage[latin9]{inputenc}
\usepackage{color}
\usepackage{babel}
\usepackage{float}
\usepackage{amsmath}
\usepackage{amssymb}
\usepackage{graphicx}
\usepackage[unicode=true,pdfusetitle,
 bookmarks=true,bookmarksnumbered=false,bookmarksopen=false,
 breaklinks=false,pdfborder={0 0 0},pdfborderstyle={},backref=false,colorlinks=false]
 {hyperref}
\hypersetup{
 colorlinks,linkcolor=blue,citecolor=blue,urlcolor=blue}

\makeatletter

\providecommand{\tabularnewline}{\\}

\makeatother

\begin{document}
\title{Skew-scattering-induced giant antidamping spin-orbit torques: Collinear
and out-of-plane Edelstein effects at two-dimensional material/ferromagnet
interfaces}
\author{Frederico Sousa}
\affiliation{Department of Physics, University of York, YO10 5DD, York, United
Kingdom}
\author{Gen Tatara }
\affiliation{RIKEN Center for Emergent Matter Science (CEMS), 2-1 Hirosawa, Wako,
Saitama 351-0198, Japan}
\author{Aires Ferreira}
\email{aires.ferreira@york.ac.uk}

\affiliation{Department of Physics, University of York, YO10 5DD, York, United
Kingdom}
\begin{abstract}
Heavy metal/ferromagnet interfaces feature emergent spin-orbit effects
absent in the bulk materials. Because of their inherent strong coupling
between spin, charge and orbital degrees of freedom, such systems
provide a platform for technologically sought-after spin-orbit torques\,(SOTs).
However, the microscopic origin of purely interfacial antidamping
SOT, especially in the ultimate atomically thin limit, has proven
elusive. Here, using two-dimensional (2D) van der Waals materials
as a testbed for interfacial phenomena, we address this problem by
means of a microscopic framework accounting for band structure effects
and impurity scattering on equal footing and nonperturbatively. A
number of unconventional and measurable effects are predicted, the
most remarkable of which is a giant enhancement of antidamping SOT
in the dilute disorder limit induced by a robust skew scattering mechanism,
which is operative in realistic interfaces and does not require magnetic
impurities. The newly unveiled skew scattering mechanism activates
rich semiclassical spin-charge conversion effects that have gone unnoticed
in the literature, including a collinear Edelstein effect with nonequilibrium
spin polarization aligned with the direction of the applied current. 
\end{abstract}
\maketitle
When a current is driven through a surface with broken inversion symmetry,
a nonequilibrium spin polarization  is induced due the spin-orbital-entangled
character of electronic wavefunctions. If coupled to a ferromagnetic
system, the emergent spin polarization transfers angular momentum
to local spin moments, changing their state by exerting a torque $\mathbf{\mathbf{T}}\propto\mathbf{m}\times\mathbf{S}$
\citep{RevModPhys.91.035004,Nature_SOT_magnetic_switching,Nature_SOT_SHE_magnetic_switching,Fast_Magnetization_Switching}.

Current-induced SOTs are conventionally classified into two broad
categories depending on their behavior under time-reversal $\mathcal{T}$:
the $\mathbf{m}$-odd or \emph{field-like} SOT that affects the precession
around the effective magnetic field and the $\mathbf{m}$-even or
\emph{antidamping }torque that renormalizes the Gilbert damping and
is responsible for the magnetization switching \citep{LLG_equation,Garello13,Titov_2DEGRashba_SOT}.
Thinning down heterointerfaces and devices by utilizing van der Waals
(vdW) crystals opens up intriguing possibilities. Fueled by the discovery
of ferromagnetism in 2D materials, recent works have reported SOT
switching of vdW-bonded ferromagnets (FMs), an important stepping-stone
towards the all-electrical control of atomically thin spin memories
\citep{2D_CrI3_Magnetic_Layer_Dependent,2D_Ferromagnetism_CrGeTe_vdW,SOT_Switching_Fe3GeTe2,SOT_Switching_Fe3GeTe2_SciAdv}.
Conversely, nonmagnetic 2D crystals with heavy atomic elements can
be used as a \emph{source }of interfacial\emph{ }SOT. Experiments
employing WTe$_{2}$ \citep{SOT2D_WTe2_17,SOT2D_WTe2_17_b,SOT2D_WeTe2_19},
a transition metal dichalcogenide (TMD) with reduced crystal symmetry,
have observed strong out-of-plane antidamping torques, which are relevant
for high-density magnetic memory applications. While these findings
represent significant steps towards SOT  devices based entirely on
2D vdW crystals~\citep{SOT_2DvdW_device_abinitio,SOT_2DvdW_graphene_device_abinitio},
the understanding of the underlying mechanisms remains in its infancy. 

In this article, we identify and\,\,quantify the dominant scattering-dependent
mechanisms of SOT for a wide class of weakly disordered 2D vdW monolayers
and their typical heterostructures.\textcolor{black}{{} To this end,
we develop a microscopic framework wherein all }interfacial spin-dependent
interactions experienced by charge carriers\textcolor{black}{{} (including
magnetic exchange interaction in arbitrary direction) are treated
nonperturbatively, which gives access to the full SOT angular dependence
so far inacessible by rigorous diagrammatic treatments. An exact ressummation
scheme of single-impurity diagrams is found to capture a unique interplay
between spin and lattice-pseudospin degrees of freedom that activates
}\textcolor{black}{\emph{all}}\textcolor{black}{{} SOT components compatible
with hexagonal symmetry \citep{PhysRevB.94.054415} i.e. $\mathbf{T}=t_{\textrm{o}1}(\phi)\,\mathbf{m}\times(\hat{z}\times\mathbf{J})+t_{\textrm{e}1}(\phi)\,\mathbf{m}\times(\mathbf{m}\times(\hat{z}\times\mathbf{J}))+t_{\textrm{o}2}(\phi)\,\mathbf{m}\times(\mathbf{m}\times\hat{z})(\mathbf{m}\cdot\mathbf{J})+t_{\textrm{e}2}(\phi)\,\mathbf{m}\times\hat{z}\,(\mathbf{m}\cdot\mathbf{J})$,
with $\hat{z}$ the versor normal to the 2D plane, $\phi=\arccos(\mathbf{m}\cdot\hat{z})$,
$\mathbf{J}$ the current density and $t_{\text{e}(\text{o})i}$ with
$i=1,2$ are torque efficiencies. This unusual proliferation of SOTs
that }\textcolor{black}{\emph{scale linearly with the conductivity}}\textcolor{black}{{}
stems from skew-scattering-induced nonequilibrium spin polarization
with  components along }\textcolor{black}{\emph{all}}\textcolor{black}{{}
spatial directions. Surprisingly, as shown below, the $\mathbf{m}$-even
torques acquire significant magnitudes already for layers with $C_{6v}$
symmetry. These technologically relevant SOTs are highly sensitive
to the impurity potential strength as well as proximity effects that
reduce the point group symmetry. This is encouraging as local symmetry
breaking and disorder landscape can be engineered with nanofabrication
methods. }

\begin{figure*}[t]
\noindent \begin{centering}
\includegraphics[width=0.9\textwidth]{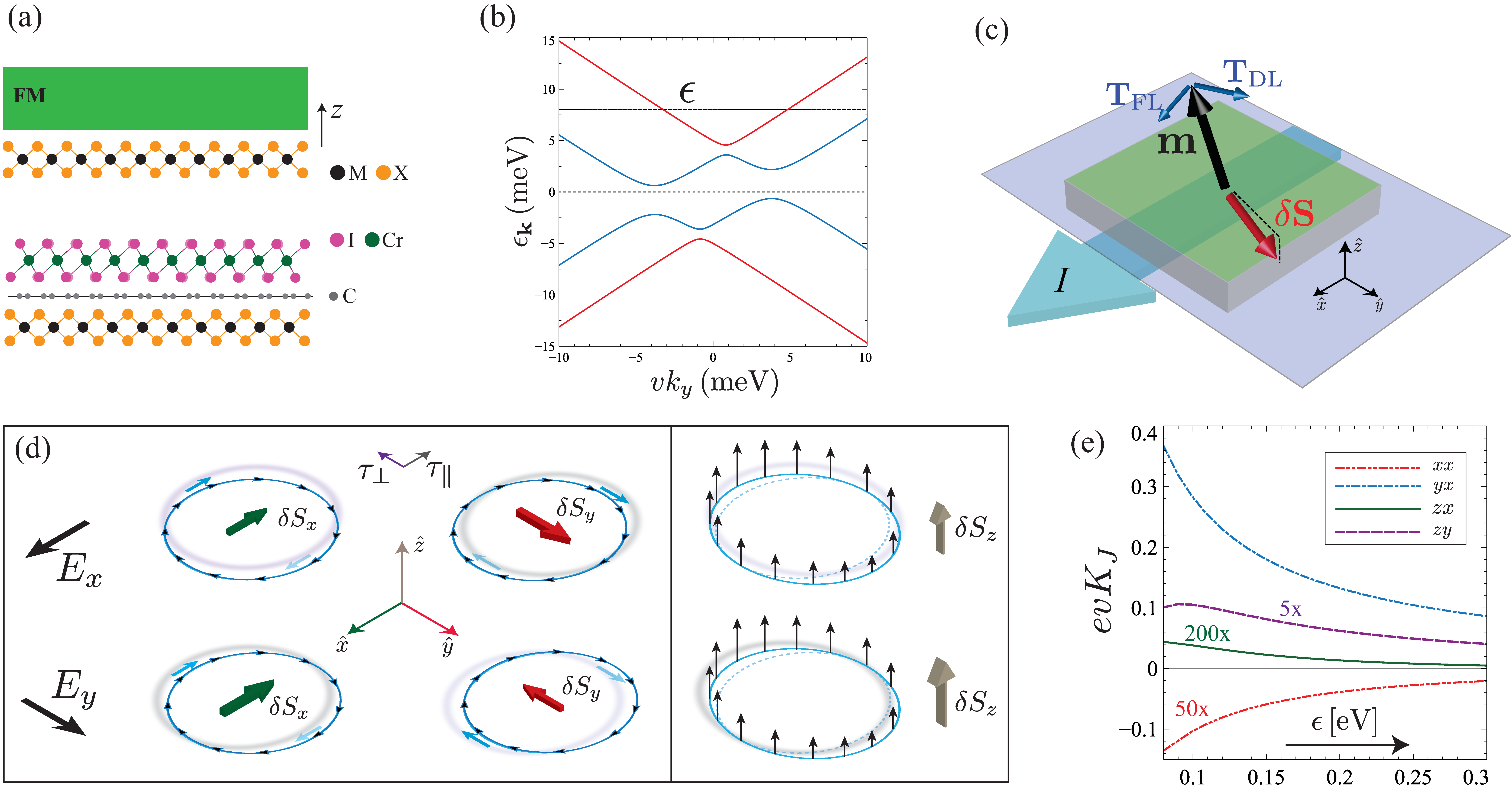}
\par\end{centering}
\noindent \caption{\label{fig:01}(a) Schematic of a TMD/thin-film-FM (top) and TMD/graphene/2D-FM
bilayer (bottom). (b) Electronic structure around $K$ points in reference
heterostructure (plotted along a path with $k_{x}=0$). (c) Geometry
of 2D material SOT-operated device. Direction of SOTs is indicated
for current applied along positive $\hat{x}$ axis assuming $\epsilon>0$.
(d) Current-induced distortion of the Fermi surface in the relaxation-time
approximation (gray) and full distortion accounting for skew scattering
events, $\delta f_{\mathbf{k}}\propto\tau_{\parallel}\,\mathbf{\hat{k}}\cdot\mathbf{E}+\tau_{\perp}\,(\mathbf{\hat{k}}\times\mathbf{E})\cdot\hat{z}$
(purple). 3D arrows depict the net non-equilibrium spin polarization
density. (e) Fermi energy dependence of SOT efficiencies for reference
$C_{6v}$-invariant monolayer system. Parameters: $\lambda=20$ meV,
$\Delta_{\textrm{xc}}=15$ meV, $\phi=\pi/11$, $n=10^{11}$ cm$^{-2}$
and $u_{0}=1.2$ eV$\cdot$nm$^{2}$.}
 
\end{figure*}
\emph{Model.}---The low-energy excitations in vdW heterostructures
made of typical monolayer compounds, such as graphene and TMDs (see
Figs. \ref{fig:01}(a)-(b)), are governed by the following generalized
Dirac-Rashba model, where $\xi=\pm$ signs refer to valleys $K(+)$
and $K^{\prime}(-)$, 

\begin{equation}
\mathcal{H}_{\xi}=\int d\mathbf{x}\,\psi_{\xi}^{\dagger}\left[v\,\boldsymbol{\Sigma}\cdot(-\imath\boldsymbol{\nabla}+\boldsymbol{\mathcal{A}}_{\xi})+\xi\Delta\Sigma_{z}+\mathcal{A}_{\xi}^{0}-\epsilon\right]\psi_{\xi},\label{eq:Hamiltonian_full-1}
\end{equation}
$(\psi_{\xi},\psi_{\xi}^{\dagger})\equiv(\psi_{\xi}(\mathbf{x}),\psi_{\xi}^{\dagger}(\mathbf{x}))$
are 4-component spinor fields defined on the internal spaces of sublattice
(``$\Sigma$'') and spin (``$s$''), $v\simeq10^{6}$ m/s is the
bare Fermi velocity of 2D Dirac fermions and $\epsilon$ is the Fermi
energy \citep{Hernando06,Kochan17,MilletariFerreira16a,MilletariFerreira16b,manuel_covariant_laws,manuel_optimalS2C}.
The gauge-field components $\mathcal{A}_{\xi}^{\mu}$ ($\mu=0,x,y,z$)
in the Hamiltonian (\ref{eq:Hamiltonian_full-1}) are $2\times2$
matrices of the form $\mathcal{A}_{\xi}^{\mu}=\sum_{a=x,y,z}\mathcal{A}_{\xi a}^{\mu}s_{a}$,
which account for all possible spin-dependent effects \citep{Tokatly08}.
The Pauli matrices $\Sigma^{a}$ and $s^{a}$ ($a=x,y,z$) all anticommute
with $\mathcal{T}$, so that their products are invariant under time-reversal
(which also interchanges valleys $\xi\leftrightarrow-\xi$). The staggered
on-site potential ($\Delta$) describes orbital-gap opening due to
broken sublattice symmetry \citep{TMD1,TMD2}. The ubiquitous interfacial
Bychkov-Rashba (BR) effect, with coupling strength $\lambda$, is
captured by the gauge-field components $\mathcal{A}_{\xi y}^{x}=-\mathcal{A}_{\xi x}^{y}=\lambda/v$
\citep{Rashba_Bychkov_SOC,Rashba09}. Other spin-orbit effects include
intrinsic spin-orbit coupling (SOC) of McClure-Yafet-Kane-Mele type
($\mathcal{A}_{\xi}^{z}=\lambda_{0}s_{z}$) and spin-valley coupling
($\mathcal{A}_{\xi}^{0}=\xi\lambda_{\textrm{sv}}s_{z}$) \citep{Kane_Mele_QSHE},
which plays a crucial role in spin relaxation \citep{Cummings17,Manuel_SpinRelax_GrapheneTMD,Ghiasi17,Benitez18}
and spin Hall effect \citep{manuel_covariant_laws}.

The interaction between the spin of 2D carriers and the local moments
in the adjacent FM layer induces an interfacial exchange field, $\mathcal{A}_{\xi}^{\textrm{0}}=-\Delta_{\textrm{xc}}\,\mathbf{m}\cdot\mathbf{s}$,
with $\Delta_{\textrm{xc}}>0$. A nonzero in-plane\textcolor{blue}{{}
}\textcolor{black}{exchange coupling} ($m_{x,y}$) lifts the rotational
symmetry of the effective Hamiltonian (\ref{eq:Hamiltonian_full-1}),
which will entail the co-existence of in-plane and out-of-plane nonequilibrium
spin-polarization (more on this later). Representative energy bands
for a reference graphene/FM heterostructure are shown in Fig.~\ref{fig:01}(b).
Without loss of generality, we choose the in-plane exchange coupling
along the $\mathbf{x}$-axis and write $\mathbf{m}=m_{x}\hat{x}+m_{z}\hat{z}\equiv\sin\phi\hat{x}+\cos\phi\hat{z}$
(Fig.~\ref{fig:01} (c)). Thus, our microscopic theory encompasses
both perpendicularly and in-plane magnetized SOT devices. The lowest
carrier density regime in Fig.\,\ref{fig:01}(b) exists only in the
anisotropic case ($m_{x}\neq0$) and exhibits an electron ($\epsilon>0$)
or hole ($\epsilon<0$) pocket away from the $K$ point. At intermediate
carrier densities, we find a ``Mexican-hat'' dispersion, followed
by a narrow spin-gap region with a single (distorted) Fermi ring \citep{AHE_Manuel_Dirac}.
At high electronic density, the two spin-split bands with counterrotating
spin textures are occupied. The spin texture of spin-majority states
(colored blue in Fig. \ref{fig:01}(b)), as well as the current-induced
distortion of the Fermi surface, are illustrated in Fig. \ref{fig:01}(d).
The out-of-plane component of the spin texture can be triggered by
an exchange field, spin-valley coupling or competition between BR
and orbital effects (see below and Appendix A for more details). 

\emph{Semiclassical picture}.---We first determine the interfacial
SOTs activated by impurity scattering mechanisms using a semiclassical
analysis, then we derive a general microscopic picture for current-induced
spin polarization in 2D vdW heterostructures and discuss its consequences.
For the first part, we restrict the discussion to $C_{6v}$-invariant
models, which already display the essential phenomenology. A general
symmetry-based analysis of the spin-charge response function is given
in Appendix B. The first step is to determine the spin texture at
the Fermi energy. Perturbation theory in the anisotropy parameter
yields, after a long but straightforward calculation, $\mathbf{s}_{\mathbf{k}\nu}=\nu(\mathbf{s}_{\mathbf{k}}^{0}+\delta\mathbf{s}_{\mathbf{k}})$,
with the signs $\nu=\pm1$ for majority/minority-spin bands (marked
blue/red in Fig.~\ref{fig:01}(b)), $\mathbf{s}_{\mathbf{k}}^{0}=\varrho_{\parallel}\,\hat{k}\times\hat{z}+m_{z}\varrho_{\perp}\hat{z}$
the noncoplanar spin texture induced by an out-of-plane exchange combined
with BR effect and $\delta\mathbf{s}_{\mathbf{k}}=(\gamma_{\parallel}+\mu_{\parallel})\,\mathbf{m}\cdot\hat{k}+2\mu_{\parallel}\,\hat{k}\times(\hat{k}\times\mathbf{m})+m_{z}\mu_{\perp}\hat{k}\times\mathbf{m}$
the correction induced by $m_{x}$, or, in a more intuitive form 
\begin{equation}
\delta\mathbf{s}_{\mathbf{k}}=m_{x}\left(\gamma_{\parallel}\hat{x}+\mu_{\parallel}\left(\cos2\theta\,\hat{x}+\sin2\theta\,\hat{y}\right)-m_{z}\mu_{\perp}\sin\theta\,\hat{z}\right),\label{eq:spin_texture}
\end{equation}
with $\theta$ the wavevector angle. In these expressions, all the
coefficients $\{\varrho_{\parallel(\perp)}$, $\gamma_{\parallel}$,
$\mu_{\parallel(\perp)}\}$ are functions of $k=|\mathbf{k}|$, $\lambda$
and $m_{z}^{2}$. The spin-helical component (in $\mathbf{s}_{\mathbf{k}}^{0}$)
yields the well-known inverse spin-galvanic effect ($\mathbf{S}\propto\varrho_{\parallel}\hat{z}\times\mathbf{J}$),
which is ubiquitous in heterointerfaces \citep{Burkov04,Garate10,Yokoyama10,Shen14,TI_SOT,manuel_optimalS2C}.
This nonequilibrium spin polarization exerts a field-like torque $\mathbf{T}_{\textrm{o}1}\propto\mathbf{m}\times\left(\hat{z}\times\mathbf{J}\right)$.
Concurrently, the $m_{x}$-induced distortion to the spin texture
{[}Eq.\,\,(\ref{eq:spin_texture}){]} produces\emph{ out-of-plane}
spin polarization $S_{z}\propto m_{x}m_{z}J_{y}$, when a current
is applied transverse to the in-plane anisotropy axis (see Fig.~\ref{fig:01}(d)).
This generates a field-like SOT that is sensitive to the direction
of the applied current, $\mathbf{T}_{\textrm{o}2}\propto\mathbf{m}\times(\mathbf{m}\times\hat{z})(\mathbf{m}\cdot\mathbf{J})$.
This newly unveiled effect, which can be traced back to the unique
Dirac-Rashba character of electronic states, still occurs when the
two spin-slit bands (with opposite-in-sign $\hat{z}$-polarizations)
are populated. This avoided cancellation of nonequilibrium out-of-plane
spin polarization stems from the interplay of pseudospin and spin
angular momentum, which renders contributions from spin-split bands
inequivalent. This differs from 2D electron gases, for which the only
robust SOT is $\mathbf{T}_{\textrm{o}1}$ \citep{Titov_2DEGRashba_SOT}. 

To explain the emergence of robust antidamping SOTs, we add the effect
of a finite transverse scattering time to the picture. Semiclassicaly,
the nonequilibrium spin polarization is obtained as $\mathbf{S}=\sum_{\mathbf{k}\nu}\mathbf{s}_{\mathbf{k}\nu}\delta f_{\mathbf{k}\nu}$,
where $\delta f_{\mathbf{k}\nu}\propto\tau_{\parallel}^{\nu}\,\hat{k}\cdot\mathbf{E}+\tau_{\perp}^{\nu}\,(\hat{k}\times\mathbf{E})_{z}$
is the deviation of the distribution function away from equilibrium
\footnote{To simplify the analysis, we neglect the $\theta$-dependence of the
transport times. This is justified since the anisotropy present in
the equilibrium spin texture suffices to capture the qualitative behavior
of the density-current response function.}. Consider an electric field applied along $\hat{\mathbf{x}}$. The
Fermi surface is shifted perpendicular to the applied current by an
amount $\delta f_{\mathbf{k}\nu}^{\perp}\propto\left(\tau_{\perp}^{\nu}\sin\theta\right)E_{x}$.
This results in an extrinsic anomalous Hall effect \citep{AHE_Manuel_Dirac},
but it also provides an efficient mechanism for current-induced collinear
spin polarization $S_{x}$ as shown here. Skew scattering plays an
essential role as there must be an imbalance between scattering cross
sections at angles $\pm\theta$, relative to $\mathbf{E}$, otherwise
all the states in the Fermi surface will have their $S_{x}$ component
cancelled by states with opposite angle. This mechanism is operative
under rather general conditions because the spin-orbit-coupled carriers
experience an average out-of-plane Zeeman field $\hat{z}\cdot\langle\mathbf{s}_{\mathbf{k}\nu}\rangle_{\textrm{FS}}\propto m_{z}$
that breaks the left/right symmetry of scattering events, regardless
of the impurity potential specifics, where $\langle...\rangle_{\textrm{FS}}$
denotes the average over the Fermi surface. After performing the angular
integration accounting for a finite $\tau_{\perp}^{\nu}$, we easily
find the magnetoelectric effect: $\mathbf{S}\propto m_{z}\varrho_{\parallel}\mathbf{E}$.
The generation of collinear nonequilibrium spin polarization can be
extremely efficient in the clean limit due to its inherent semiclassical
scaling $\tau_{\perp}\propto\tau_{\parallel}\propto n^{-1}$ (where
$n$ is the impurity density) \footnote{This phenomenon is distinct from quantized magnetoelectric effect
inside the surface gap of topological insulator/ferromagnet interfaces,
$\mathbf{S}=-\sigma_{H}\mathbf{E}$ with $\sigma_{H}$ the Hall conductance,
which is a topological effect \citep{Garate10}.}. This phenomenon, which we term \emph{collinear Edelstein effect,}
contributes with an antidamping SOT $\mathbf{T}_{\textrm{e}1}\propto\mathbf{m}\times(\mathbf{m}\times(\hat{z}\times\mathbf{J}))$.
From Eq.~(\ref{eq:spin_texture}), one can easily conclude that the
skewness also activates an out-of-plane spin response, $S^{z}\propto\tau_{\perp}m_{x}m_{z}E_{x}\propto m_{x}m_{z}^{2}E_{x}$.
This yields an antidamping torque $\mathbf{T}_{\textrm{e2}}\propto\mathbf{m}\times\hat{z}\,(\mathbf{m}\cdot\mathbf{J})$.
These novel SOTs, which scale favorably with the conductivity $\text{\ensuremath{\sigma_{0}\propto}}\,\epsilon\,\tau_{\parallel}/\hbar\gg1$,
are our central result. The semiclassical mechanisms are summarized
in Fig. \ref{fig:01}(d).

\begin{figure}[t]
\centering{}\includegraphics[width=0.9\columnwidth]{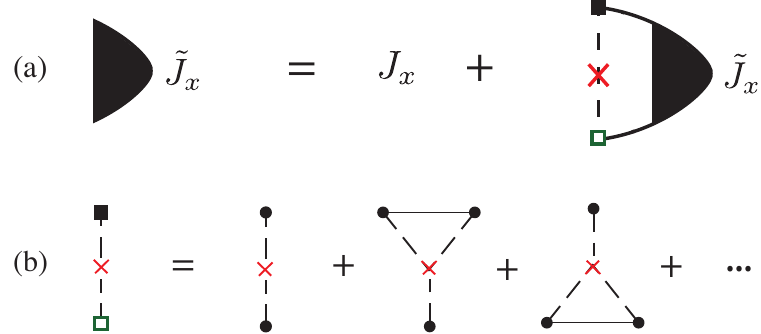}\caption{\label{fig:02}Diagrammatic expansion of the response function. (a)
Bethe-Salpeter equation for the charge current vertex in the $R$--$A$
sector. (b) Skeleton expansion of the $T$-matrix ladder. Full (open)
square denotes a $T$ ($T^{\dagger}$) matrix insertion, while circles
represent electron--impurity interaction vertices. The red $\times$
stands for impurity density insertion ($n$).}
\end{figure}

\emph{T-matrix diagrammatic approach}.---To derive an accurate microscopic
theory of SOT that includes intrinsic effects and disorder corrections
(impurity scattering) selfconsistently, we extend the controlled diagrammatic
technique developed in Refs. \citep{MilletariFerreira16a,MilletariFerreira16b,manuel_covariant_laws,manuel_optimalS2C}
to arbitrary multi-band models. Our approach has two essential features.
First, it is fully nonperturbative in the energy scales of the bare
Hamiltonian, which includes orbital mass $\Delta$, exchange field
vector, BR interaction $\lambda$ and other couplings. This technique
allow us to explore rich scenarios, including the experimentally relevant
regime of proximitized materials with competing energy scales e.g.,
$\lambda\approx\Delta_{\textrm{xc}}\approx\epsilon$. Simple analytical
expressions can be obtained to leading order in $m_{x}$ by developing
the Green's functions in Dyson series \citep{Titov_2DEGRashba_SOT}.
Second, the three-leg spin-charge correlation or vertex function $\Gamma_{i\alpha\beta}(\mathbf{x},\mathbf{y},\mathbf{z})=\langle T\,J_{i}(\mathbf{x})\Psi_{\alpha}(\mathbf{y})\Psi_{\beta}^{\dagger}(\mathbf{z})\rangle$
is evaluated by re-summing \emph{all} single-impurity Feynman diagrams,
which provides the dominant contribution to the spin-charge response
functions in the dilute impurity regime. This is accomplished by writing
a Bethe-Salpeter equation with $T$ matrix insertions \citep{MilletariFerreira16a},
which is more general and accurate than the standard approach based
on ladder diagrams (Figs.~\ref{fig:02}(a)-(b)). This allow us to
obtain virtually exact results and explore the crossover between the
standard weak Gaussian limit and the important unitary scattering
regime, which physically corresponds to resonant scattering from vacancies
or adatoms \citep{RS_graphene,Ferreira11rs}.

We are interested in SOTs generated by weakly disordered 2D materials
and thus focus our subsequent analysis on Fermi surface processes.
The latter are captured by the spin density--charge current response
function \citep{AHE_Kubo_Streda,MilletariFerreira16a}
\begin{equation}
K_{ai}=\frac{1}{2\pi}\,\text{Tr}\left[s_{a}\left\langle G^{+}J_{i}G^{-}\right\rangle \right],\label{eq:Kubo-Streda}
\end{equation}
with $G^{\pm}$ the retarded(+)/advanced(-) Green's function, $J_{i}=-e\,\partial_{p_{i}}\mathcal{H}_{\mathbf{p}}=-ev\Sigma_{i}$
the charge current operator and $\textrm{Tr}$ the trace over all
degrees of freedom. Here the angular brackets denote disorder averaging
and $\mathcal{H}_{\mathbf{p}}$ is the extension of the single-particle
Hamiltonian to the space of two valleys. Since our aim is to develop
a generic SOT theory, which does not rely on the existence of spinful
scattering centers, such as spin-orbit-active impurities \citep{Pesin12,Federov13,Ferreira14,Pachoud14,Huang14},
we assume a standard scalar short-range potential $V(\mathbf{x})=u_{0}\sum_{i=1}^{N}\delta(\mathbf{x}-\mathbf{x}_{i})$,
where $\mathbf{x}_{i}$ are random impurity locations and $u_{0}$
parametrizes the potential scattering strength. Leading terms $K_{ai}\propto1/n$
in the dilute impurity regime are obtained by replacing in Eq. (\ref{eq:Kubo-Streda})
$\langle G^{+}J_{i}G^{-}\rangle\rightarrow\mathcal{G}_{\mathbf{p}}^{+}\tilde{J_{i}}\mathcal{G}_{\mathbf{p}}^{-}$,
where $\mathcal{G}_{\mathbf{p}}^{\pm}$ is the disorder-averaged Green's
function and $\tilde{J_{i}}$ is the renormalized vertex (Figs. \ref{fig:02}(a)-(b)).
The final trace in Eq. (\ref{eq:Kubo-Streda}) is carried out using
an exact SO(5) decomposition of the response function; technical details
will be published elsewhere \footnote{A. Veneri, F. Sousa, and A. Ferreira (to be published).}. 

\emph{Results: graphene}-\emph{based heterostructures}.---Armed with
this formalism, we evaluate the SOTs and determine their efficiency.
Within linear response theory, we write $\mathbf{\mathbf{T}}=d^{-1}\,\mathbf{m}\times\mathbf{H}_{T}$,
where $d$ is the FM thin film thickness $\mathbf{H}_{T}=-\Delta_{\textrm{xc}}\,\hat{K}^{J}\cdot\mathbf{J}$
is the current-induced spin-orbit field and $\hat{K}^{J}\equiv\hat{K}\cdot\hat{\sigma}^{-1}$,
with $\hat{\sigma}$ is the conductivity tensor, is a $3\times2$
matrix that quantifies the underlying SOC transport effects. The earlier
semiclassical picture suggests the decomposition (to leading order
in $m_{x}$)
\begin{equation}
K^{J}=\begin{pmatrix}m_{z}\kappa_{\parallel}^{ss} & \kappa_{E}\\
-\kappa_{E} & m_{z}\kappa_{\parallel}^{ss}\\
m_{x}\kappa_{zx}^{ss} & m_{z}m_{x}\kappa_{zy}
\end{pmatrix},\label{eq:Ktensor}
\end{equation}
where the superscript $ss$ marks the responses activated by skew
scattering. The Fermi energy dependence of $K^{J}$ for a graphene
heterostructure is shown in Fig.~\ref{fig:01}(e). The highly efficient
Edelstein-type response ($\kappa_{E}\sim0.4$ for $\epsilon\sim0.1$
eV) is reminiscent of  topological surface states and nonmagnetic
graphene/TMD bilayers \citep{TI_SOT,manuel_optimalS2C}. This process
is accompanied by the generation of robust out-of-plane spin polarization.
This is at variance with 2D electron gases in Rashba ferromagnets,
for which $K_{zy}^{J}\rightarrow0$ in the weak scattering limit \citep{Titov_2DEGRashba_SOT}.
Concurrently, the newly unveiled skew scattering mechanism, which
is operative in all systems with $m_{z}\neq0$, enriches the class
of SOTs to include $\mathcal{T}$-odd ($\mathbf{m}$-even) terms.
Despite the moderate scattering potential strength in Fig.~\ref{fig:01}(e),
a collinear Edelstein response is induced $K_{ii}^{J}$ ($i=x,y$).
The total spin-orbit field thus comprises $\mathbf{H}_{T}^{\textrm{e}1}\propto\mathbf{m}\times\mathbf{J}\times\hat{z}$
and $\mathbf{H}_{T}^{\textrm{e}2}\propto\hat{z}\,\mathbf{m}\cdot\mathbf{J}$
antidamping contributions. Owing to its skew-scattering origin, the
spin-orbit fields scale \emph{linearly} with the conductivity with
an efficiency $K_{ii}^{J}\sim(\sigma_{0})^{0}\sim n^{0}$. This behavior
is notoriously different from  predicted $\mathcal{T}$-odd torques
for topological insulators, whose quantum-side-jump origin \citep{MilletariFerreira16b}
yields $K_{ii}^{J}\sim1/\sigma_{0}$ in the clean limit \citep{TI_SOT}.

\begin{figure}[t]
\centering{}\includegraphics[width=1\columnwidth]{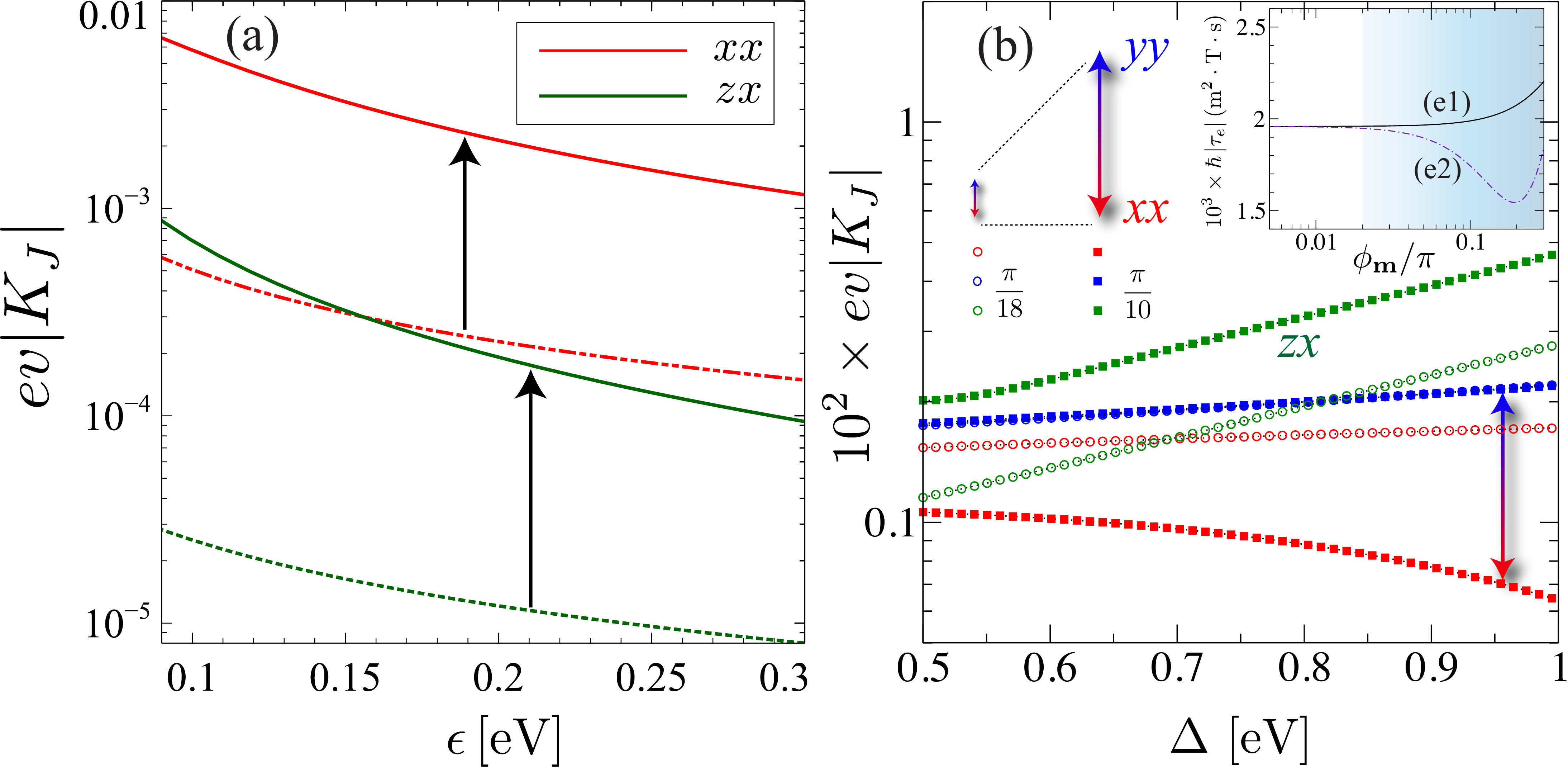}\caption{\label{fig:03}(a) Giant enhancement of damping-like SOT due to skew
scattering for proximitized graphene. Dashed (solid) lines are calculated
in the weak (unitary) scattering regime with $u_{0}=0.1$ eV$\cdot$nm$^{2}$($u_{0}\rightarrow\infty$).
Other parameters as in Fig. \ref{fig:01}(e). (b) Highly anisotropic
SOT generated by TMDs in the unitary limit. Orbital-gap-dependence
of $\mathcal{T}$-odd SOTs is evaluated at fixed carrier density $n_{e}\simeq4.7\times10^{13}$
cm$^{-2}$ for $\phi_{\mathbf{m}}=\pi/10$ (squares) and $\phi_{\mathbf{m}}=\pi/18$
(circles). Inset: angle dependence of $\tau_{\textrm{e}1(2)}$ for
$\Delta=0.75$ eV. Shaded area indicates nonperturbative region where
high-order harmonics in $t_{ei}(\phi)$  become prominent. Other parameters:
$\Delta_{\textrm{xc}}=0.1$ eV, $\lambda=60$ meV, $\lambda_{\textrm{sv}}=$
3 meV and impurity density $n=10^{11}$ cm$^{-2}$.}
\end{figure}

An unprecedented sensitivity of the SOT efficiency to the potential
scattering strength is borne out by our theory. In contrast to the
Edelstein efficiency ($\kappa_{E}$), which receives slow (logarithmic)
disorder corrections \citep{manuel_optimalS2C,manuel_MDPI}, \emph{all
damping-like efficiencies} exhibit a monotonic increase with $u_{0}$.
This important feature is illustrated in Fig.~\ref{fig:03}(a), where
a ten-fold increase in both $K_{ii}^{J}$ and $K_{zx}^{J}$ approaching
the unitary regime of a resonant scatterer ($u_{0}\rightarrow\infty$)
can be observed. In the weak scattering regime ($u_{0}\rho\ll1$)
with $\epsilon\gg\{\lambda,\Delta_{\textrm{xc}}\}$, where $\rho$
is the clean density of states, the leading-order coefficients in
the $\mathbf{m}$-expansion of the current-induced torque ($t_{\textrm{e(o)}i}\equiv d^{-1}\Delta_{\textrm{xc}}\tau_{\textrm{e(o)}i}$)
admit a compact analytic form (to leading order in $m_{x}$)
\begin{align}
 & \tau_{\textrm{o}1}\simeq2\lambda^{3}\,/f_{\epsilon}\,,\quad\quad\tau_{\textrm{o}2}\simeq2\Delta_{\textrm{xc}}^{2}\lambda/f_{\epsilon}\,,\label{eq:t_odd_12}\\
 & \tau_{\textrm{e}1}\simeq u_{0}\Delta_{\textrm{xc}}\epsilon\,\lambda^{5}/(vf_{\epsilon}^{2})\,,\quad\tau_{\textrm{e}2}\simeq-\tau_{\textrm{e}1}\,,\label{eq:t_even_12}
\end{align}
where $f_{\epsilon}=v\epsilon(\lambda^{2}+\Delta_{\textrm{xc}}^{2}m_{z}^{2})$
(see Appendix C). For interpreting these results, it is important
to note that the BR coupling should be not too small compared to $k_{B}T$
so that the torques are appreciable in realistic conditions. Notably,
the slow algebraic decay with the Fermi energy $\propto\epsilon^{-1}$
in Eqs.~(\ref{eq:t_odd_12})-(\ref{eq:t_even_12}) effectively quenches
the effect of thermal fluctuations \citep{manuel_optimalS2C}, which
in principle allows room-temperature SOT operation even for samples
with weak BR effect $\lambda\approx1$ meV. Recent observations of
gate-tunable and reversible spin galvanic effect in graphene-based
vdW heterostructures at room temperature \citep{SGE_2D_Ghiasi,SGE_2D_Benitez,SGE_2D_Lin20,K2020}\textcolor{blue}{{}
}provide extra confidence that the interfacial SOTs unveiled here
can be demonstrated experimentally. 

\emph{Results: group VI dichalcogenides}---Next, we consider models
with broken sublattice symmetry ($C_{6v}\rightarrow C_{3v}$); two
examples are shown in Fig.\,\ref{fig:01}(a). As a case study, we
focus here on semiconducting TMDs, for which interfacial magnetic
exchange coupling can be up to 100 times greater than in graphene-based
heterostructures \citep{TMD_giantMEC1,TMDgiantMEC2,TMDgiantMEC3,TMDgiantMEC4}.
The presence of a orbital-gap in TMDs ($E_{g}=2\Delta$) modifies
the $\mathbf{k}$-space spin texture dramatically. We find that the
``orbital mass'' ($\Delta\gg\lambda$) stabilizes a giant equilibrium
out-of-plane spin polarization around $K$ points even in the absence
of spin-valley coupling (see Appendix A). To determine the antidamping
spin-orbit fields $\mathbf{H}_{T}^{\textrm{e}1(2)}$, we evaluate
the spin-charge correlation vertex of the full model {[}Eq.(\ref{eq:Hamiltonian_full-1}){]}.
Figure \ref{fig:03}b shows the SOT evolution with the orbital gap.
Its most salient feature is a strong enhancement of out-of-plane antidamping
efficiency ($K_{zx}$). This phenomenon is accompanied by a sizeable
collinear Edelstein effect, with tunable degree of anisotropy $\partial_{\Delta}|K_{xx}-K_{yy}|>0$
(e.g., by applying strain), as indicated by the colored arrows. The
figure of merit ($K^{J}$-efficiency) for antidamping SOT generated
by electrons in the doped regime ($\epsilon>\Delta$) reaches 4\%
with $\tau_{\textrm{e}}/\tau_{\textrm{o}}$ ratios of order $0.1$.\textcolor{black}{{}
A remark is in order. We have thus far employed the commonly used
terminology of antidamping- and field-like torques for $t_{\text{e}i}$
and $t_{\text{o}i}$, respectively. Rigorously, one needs to expand
these terms in vector spherical harmonics to truly discriminate damping
and fieldlike components \citep{PhysRevB.101.020407}, especially
when considering strong in-plane magnetization. In that case $t_{\text{e}i}$
will yield fieldlike contributions and $t_{\text{o}i}$ dampinglike.
The leading contributions are nonetheless of the nature we have denoted
them. (The exceptions are $t_{\text{e}2}$ and $t_{\text{o}2}$ which
already at leading order are a mixture of field- and damping-like
SOTs.) This implies that the presence of $K_{zy}$, already at the
Gaussian level as unveiled here, represents a robust extrinsic source
of genuine dampinglike SOT.}

We briefly discuss the implications of our findings for the magnetization
dynamics. Crucially, as soon as a finite in-plane magnetization $m_{x}$
is included nonperturbatively in the microscopic treatment (this is
possible only within a numerical approach), the SOT efficiency tensor
$K_{J}$ acquires higher-order harmonics, which invalidates a simplistic
analysis in terms of constant ``torkances'' {[}Eqs. (\ref{eq:t_odd_12})-(\ref{eq:t_even_12}){]}
as emphasized recently in Ref. \citep{PhysRevB.102.014401}. The shortcoming
of the standard approximation can be clearly seen in the TMD-based
system, where the SOT becomes highly anisotropic {[}i.e. $\delta\tau_{12}\equiv||\tau_{\textrm{e}2}|-|\tau_{\textrm{e}1}||/|\tau_{\textrm{e}1}|\approx0.25${]}
for $\phi\approx\pi/10$ (Fig. \ref{fig:03}b), reflecting the $O$(2)
rotational symmetry breaking. \textcolor{black}{Thus, a fully-fledged
microscopic treatment for $\mathbf{T}(\phi)$ becomes indispensable
to faithfully capture the ensuing magnetization dynamics when solving
the Landau--Lifshitz--Gilbert equation \citep{RevModPhys.91.035004},
irrespective of the FM geometry and its initial macrospin configuration
$\mathbf{m}(t=0)$. The full angular dependence of the torkances is
}shown in Appendix D for a graphene-based heterostructure\textcolor{black}{.
Finally, we emphasize that the angular dependence of the torkances,
as well as the leading scaling of antidamping terms $t_{\textrm{e}1(2)}\propto(\sigma_{0})^{0}$
(resulting in $\mathbf{T}_{\textrm{e1(2)}}\propto\sigma_{0}\propto\tau_{\parallel}$)
reported here, are not captured within perturbative Kubo-Streda calculations
for disordered interfaces employed in previous works \citep{Barnas_SOT_Graphene_PRB2015,TI_SOT,Titov_2DEGRashba_SOT,PhysRevLett.108.117201}. }

\emph{Conclusions}.---In summary, we have reported a microscopic
theory of SOT generated by 2D materials proximity coupled to a ferromagnet.
The SOTs are evaluated in linear response theory for a generalized
Dirac-Rashba model describing 2D vdW heterostructures with $C_{6v}$
or $C_{3v}$ point group symmetry in the presence of smooth magnetic
textures, which is readily applicable to SOT devices with both in-plane
and perpendicular magnetization. The microscopic calculations are
carried out within a $T$-matrix diagrammatic approach that captures
the extrinsic skew scattering contribution to the current-induced
SOT inacessible by previous perturbative treatments. We find that
skew scattering from non-magnetic impurities enables the robust generation
of non-equilibrium non-coplanar spin polarization. Through a complementary
semiclassical analysis, we attribute the interfacial skew scattering
mechanism to the tilting of the Rashba spin texture caused by the
exchange coupling to the ferromagnet. The skewscattering-induced collinear
and out-of-plane inverse spin galvanic effects were shown to activate
all m-even SOTs compatible with hexagonal symmetry. Interestingly,
such m-even SOTs are massively enhanced in the resonant scattering
regime of strong impurity potentials (e.g. due to atomically sharp
defects), with ratios $|\mathbf{T}|_{\textrm{even}}/|\mathbf{T}|_{\textrm{odd}}$
on the order of 0.1 for TMD-based heterostructures with large magnetic
proximity effect. These semiclassical SOTs scale linearly with the
2D charge conductivity and thus are expected to dominate in the dilute
disorder limit. These findings put the spotlight on skew scattering
as a promising extrinsic source of technologically relevant antidamping
SOTs in weakly disordered interfaces. 

\emph{Acknowledgements}.---A.F. gratefully acknowledges the financial
support from the Royal Society through a Royal Society University
Research Fellowship. We thank K. D. Belashchenko for drawing our attention
to the vector spherical harmonic decomposition developed in Ref. \citep{PhysRevB.101.020407}. 

\section*{Appendix A: Electronic structure and pseudospin-spin texture}

\begin{figure}[b]
\includegraphics[width=0.9\columnwidth]{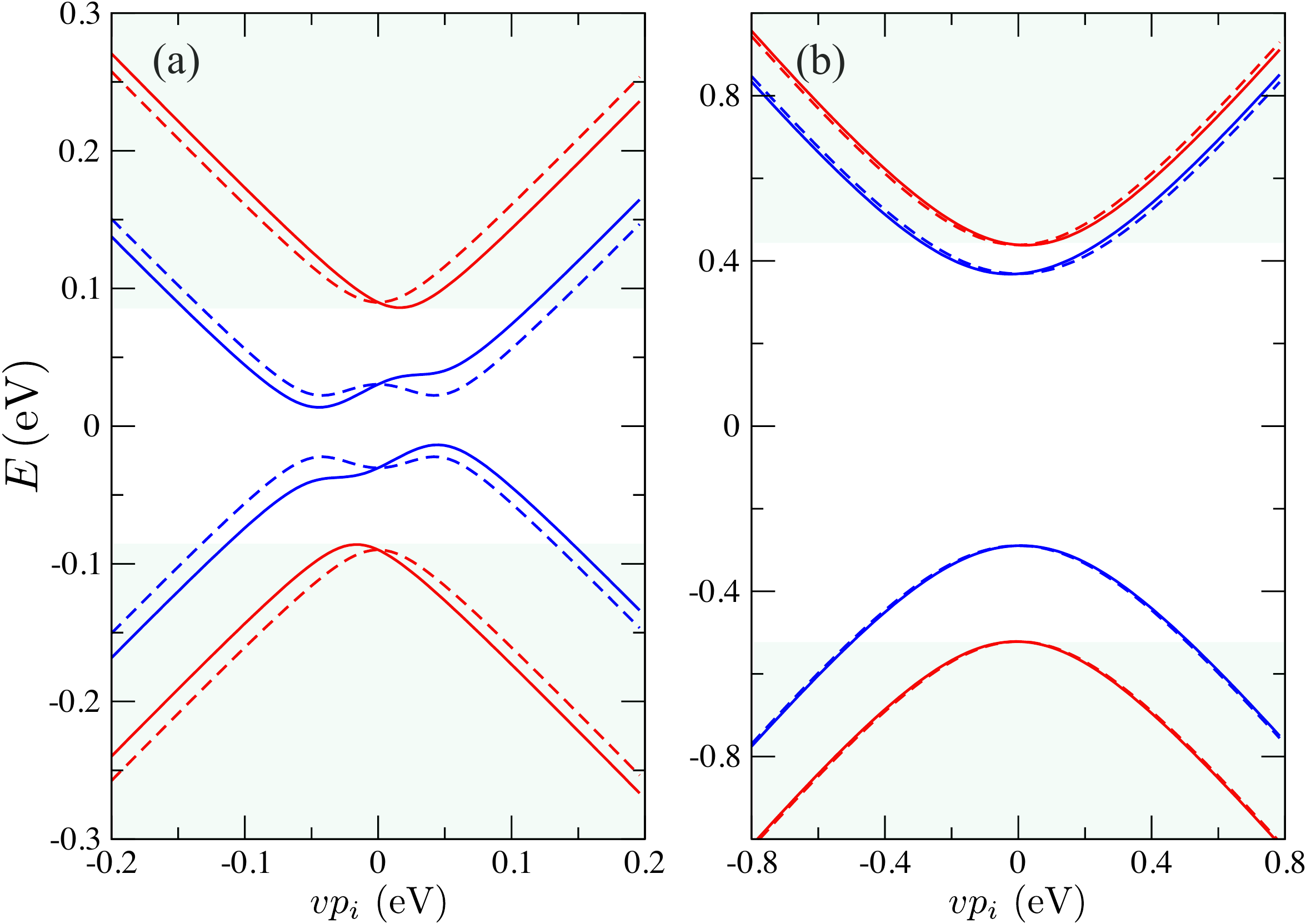}

\caption{\label{fig:spectrum}Electronic structure of a TMD|graphene|FM (a)
and TMD|FM (b) heterostructure plotted along $\mathbf{p}=(0,p_{y})^{\textrm{t}}$
(full line) and $\mathbf{p}=(p_{x},0)^{\textrm{t}}$ (dashed lines)
directions. Shaded areas highlight the spectral region II. Parameters:
for (a-b) $\lambda=40$ meV, $\mathfrak{m}_{x}=20$ meV, $\mathfrak{m}_{z}=30$
meV and for (b) $\Delta=0.4$ eV, $\lambda_{\textrm{sv}}(\epsilon<0)=150$
meV and $\lambda_{\textrm{sv}}(\epsilon>0)=5$ meV.}
\end{figure}

The low-energy Hamiltonian of $C_{3v}$-invariant vdW monolayers reads
as
\begin{equation}
H_{\xi}=\Sigma_{\mu}\,\left(vp+\mathcal{A}_{\textrm{SO}}+\mathcal{A}_{\textrm{xc}}+\mathcal{A}_{\textrm{orb}}\right)^{\mu}\,,\label{eq:1}
\end{equation}
where $p^{\mu}\equiv(-\epsilon/v,p_{x},p_{y})$ is the 3-momentum
of the interface, $\Sigma_{\mu}=(\Sigma_{0},\vec{\Sigma})$ and 
\begin{align}
 & \mathcal{A}_{\textrm{SO}}=\lambda\,(s_{y}\hat{x}-s_{x}\hat{y})+\lambda_{\textrm{KM}}s_{z}\hat{z}+\xi\lambda_{\textrm{sv}}s_{z}\hat{t}\,,\label{eq:A_SO}\\
 & \mathcal{A}_{\textrm{ex}}+\mathcal{A}_{\textrm{orb}}=(\boldsymbol{\mathfrak{m}}\cdot\mathbf{s})\hat{t}+\xi\Delta\,\hat{z}\,,\label{eq:A_2}
\end{align}
are non-Abelian gauge fields capturing all symmetry-allowed SOCs {[}Eq.
(\ref{eq:A_SO}){]}, on-site staggered potential and interfacial exchange
coupling {[}Eq. (\ref{eq:A_2}){]}. Here, $\boldsymbol{\mathfrak{m}}=-\Delta_{\textrm{xc}}\,\mathbf{m}\equiv\mathfrak{m}(\sin\phi\,\hat{\mathbf{x}}+\cos\phi\,\hat{\mathbf{z}})$
with $\pi\ge\phi\ge0$ parametrizes the exchange field. For brevity,
in this supplementary information, the analytical expressions are
provided for the strong SOC regime with $\lambda>\mathfrak{\tilde{m}}_{z}\gg\mathfrak{m}_{x}$
and $\lambda_{\textrm{KM}}=0$, where $\mathfrak{\tilde{m}}_{z}\equiv\mathfrak{m}_{z}+\xi\lambda_{\textrm{sv}}$.
Figure \ref{fig:spectrum} shows the low-energy spectrum for two representative
systems: (a) TMD/graphene/FM and (b) TMD/FM. In panel (a) only low-energy
states (within the TMD gap) are shown. In-plane magnetization ($\mathfrak{m}_{x}$)
breaks the $C_{v\infty}$ symmetry of the continuum model, rending
the Fermi surface anisotropic.

The electronic structure comprises four distinct spectral regions:

\bigskip{}

\begin{figure}
\includegraphics[width=0.9\columnwidth]{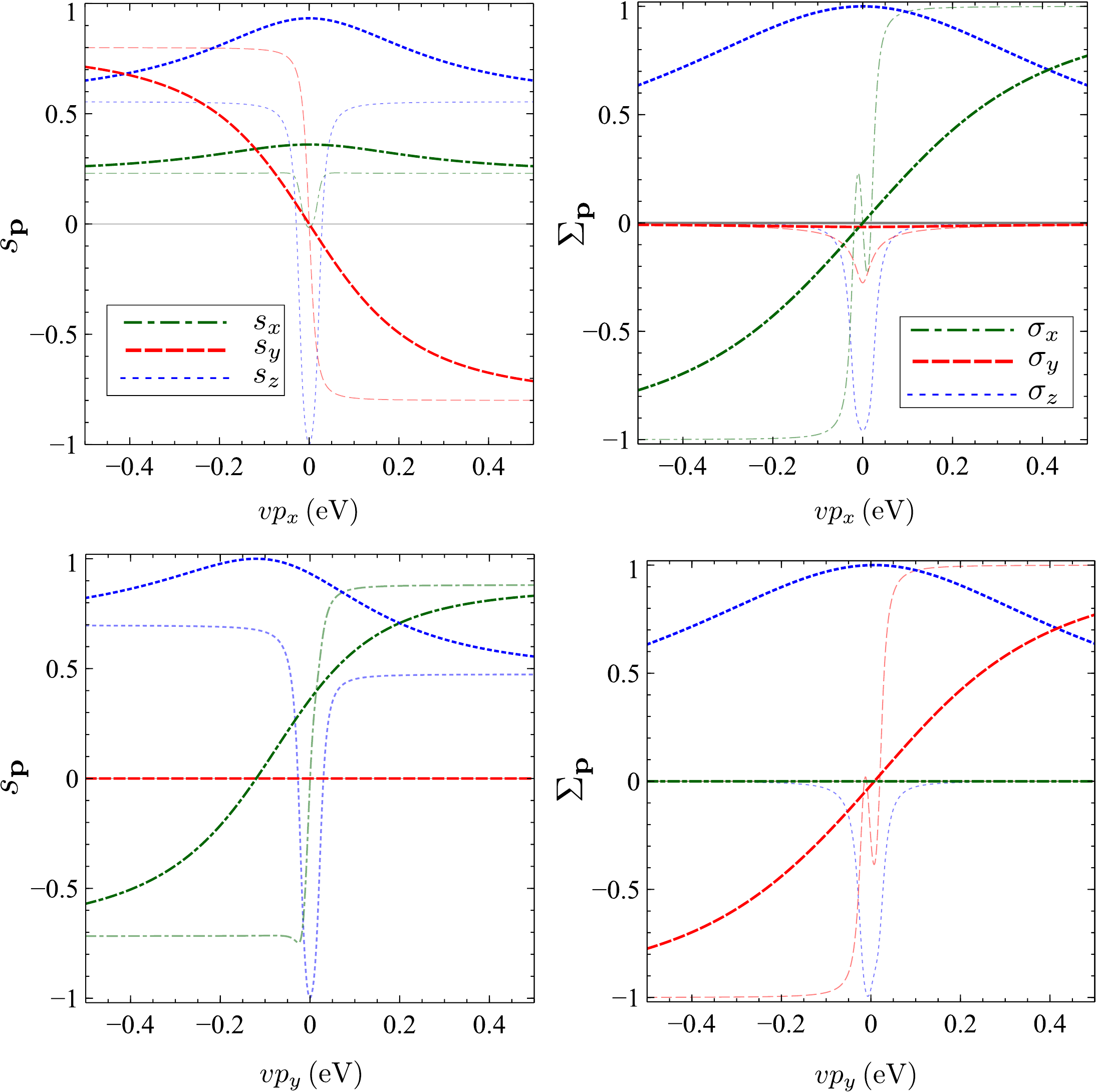}\caption{\label{fig:spin_text}Emergent pseudospin-spin texture of spin-majority
band for a 2D material|FM interface in the presence (thick lines)
and absence (thin lines) of orbital gap. The pseudospin-spin textures
are plotted along path $\mathbf{k}=(k_{x},0)$ (top panels) and $\mathbf{k}=(0,k_{y})$
(bottom panels). Band structure parameters: $\lambda=20$ meV, $\Delta=400$
meV, $\Delta_{\textrm{xc}}=15$meV (with $\phi=\pi/8$) and $\lambda_{\textrm{sv}}=0$.}
\end{figure}

\begin{itemize}
\item Regime Ia: Low energy regime where the Fermi level crosses an electron/hole
pocket for 
\begin{equation}
\epsilon_{\text{Ia}}<|\epsilon|<\epsilon_{\text{\text{Ib}}};
\end{equation}
\item Regime Ib: Very narrow energy range where the Fermi level crosses
two different Fermi rings both belonging to the spin majority band.
This happens for
\begin{equation}
\epsilon_{\text{\text{Ib}}}<|\epsilon|<\epsilon_{\text{\text{Ic}}};
\end{equation}
\item Regime Ic: Intermediate regime where the Fermi level crosses only
the spin majority band, hinting at stronger spin density responses
for 
\begin{equation}
\epsilon_{\text{Ic}}<|\epsilon|<\epsilon_{\text{II}};
\end{equation}
\item Regime II: Typical high-electronic density regime in the experiments.
Here, we have $\epsilon>\epsilon_{\text{II}}$ and the Fermi level
crosses two Fermi rings with opposite spin textures;
\end{itemize}
The expressions for the different limits are given in Table \ref{tab:limits}.
In regime II, the spin operators have the following equilibrium average
values at the Fermi energy, in the asymptotic limit $\epsilon\gg\mathfrak{m}_{z}\equiv\tilde{\mathfrak{m}}_{z}\gg\mathfrak{m}_{x}$
(here, $\theta$ is the wavevector angle with respect to $\hat{x}$
axis):
\begin{align}
\left\langle s_{x}\right\rangle  & \cong\frac{\lambda}{\sqrt{\lambda^{2}+\mathfrak{m}_{z}^{2}}}\left(1-\frac{\left(\Delta\mathfrak{m}_{z}+\lambda^{2}\right){}^{2}}{2\epsilon^{2}\left(\lambda^{2}+\mathfrak{m}_{z}^{2}\right)}\right)\sin\theta\nonumber \\
 & +\frac{\mathfrak{m}_{x}}{2\sqrt{\lambda^{2}+\mathfrak{m}_{z}^{2}}}\left(1+\frac{\mathfrak{m}_{z}^{2}+\lambda^{2}\cos2\theta}{\lambda^{2}+\mathfrak{m}_{z}^{2}}\right),
\end{align}

\begin{align}
\left\langle s_{y}\right\rangle  & \cong-\frac{\lambda}{\sqrt{\lambda^{2}+\mathfrak{m}_{z}^{2}}}\left(1-\frac{\left(\Delta\mathfrak{m}_{z}+\lambda^{2}\right){}^{2}}{2\epsilon^{2}\left(\lambda^{2}+\mathfrak{m}_{z}^{2}\right)}\right)\cos\theta\nonumber \\
 & +\frac{\mathfrak{m}_{x}}{2\sqrt{\lambda^{2}+\mathfrak{m}_{z}^{2}}}\frac{\lambda^{2}\sin2\theta}{\lambda^{2}+\mathfrak{m}_{z}^{2}},
\end{align}

\begin{align}
\left\langle s_{z}\right\rangle  & \cong\frac{\mathfrak{m}_{z}}{\sqrt{\lambda^{2}+\mathfrak{m}_{z}^{2}}}\left(1+\frac{\lambda^{2}}{2\epsilon^{2}}\frac{\Delta^{2}-\mathfrak{m}_{z}^{2}}{\lambda^{2}+\mathfrak{m}_{z}^{2}}-\frac{\mathfrak{m}_{x}\lambda\sin\theta}{\lambda^{2}+\mathfrak{m}_{z}^{2}}\right)\nonumber \\
 & +\frac{\lambda^{4}}{\epsilon^{2}}\frac{\Delta-\mathfrak{m}_{z}}{\left(\lambda^{2}+\mathfrak{m}_{z}^{2}\right){}^{3/2}},
\end{align}
for the spin majority band, the other band has opposite polarity.
The pseudospin texture, on the other hand, is:

\begin{equation}
\left\langle \Sigma_{x}\right\rangle \cong\left(1-\frac{\lambda^{2}+\Delta^{2}}{2\epsilon^{2}}\right)\cos\theta\pm\frac{\lambda\mathfrak{m}_{x}}{2\epsilon}\frac{\sin2\theta}{\sqrt{\lambda^{2}+\mathfrak{m}_{z}^{2}}},
\end{equation}
\begin{equation}
\left\langle \Sigma_{y}\right\rangle \cong\left(1-\frac{\lambda^{2}+\Delta^{2}}{2\epsilon^{2}}\right)\sin\theta\mp\frac{\alpha\mathfrak{m}_{x}}{2\epsilon}\frac{1+\cos2\theta}{\sqrt{\lambda^{2}+\mathfrak{m}_{z}^{2}}},
\end{equation}

\begin{equation}
\left\langle \Sigma_{z}\right\rangle \cong\frac{\Delta}{\epsilon}\mp\frac{\mathfrak{m}_{z}}{\sqrt{\lambda^{2}+\mathfrak{m}_{z}^{2}}}\frac{\Delta\mathfrak{m}_{z}+\lambda^{2}}{\epsilon^{2}}\left(1-\frac{\lambda\mathfrak{m}_{x}\sin\theta}{\lambda^{2}+\mathfrak{m}_{z}^{2}}\right),
\end{equation}
for the spin majority ($+$)/minority ($-$) bands. Figure (\ref{fig:spin_text})
shows the spin and pseudospin profiles along de $\hat{x}$ and $\hat{y}$
directions in momentum space. The orbital mass ($\Delta$) broadens
up the $\mathbf{p}$-space spin texture dramatically, which boosts
the generation of out-of-plane spin polarization in applied current.

\begin{table*}[t]
\begin{tabular}{|c|c|c|c|c|}
\hline 
 & $\epsilon_{\text{II}}$ & $\epsilon_{\text{\text{Ic}}}$ & $\epsilon_{\text{\text{Ib}}}$ & $\epsilon_{\text{Ia}}$\tabularnewline
\hline 
\hline 
$\,$$\,$$\Delta=0$$\,$$\,$ & $\sqrt{\mathfrak{\tilde{m}}_{z}^{2}+4\lambda^{2}}$ & $\mathfrak{\tilde{m}}_{z}+\frac{\mathfrak{m}_{x}^{2}}{2\mathfrak{\tilde{m}}_{z}}$ & $\frac{\mathfrak{\tilde{m}}_{z}\lambda}{\sqrt{\mathfrak{\tilde{m}}_{z}^{2}+\lambda^{2}}}+\frac{\lambda\mathfrak{m}_{x}\mathfrak{\tilde{m}}_{z}\sqrt{2\lambda^{2}+\mathfrak{\tilde{m}}_{z}^{2}}}{\left(\lambda^{2}+\mathfrak{\tilde{m}}_{z}^{2}\right)^{3/2}}$ & $\frac{\tilde{\mathfrak{m}}_{z}\lambda}{\sqrt{\mathfrak{\tilde{m}}_{z}^{2}+\lambda^{2}}}-\frac{\lambda\mathfrak{m}_{x}\mathfrak{\tilde{m}}_{z}\sqrt{2\lambda^{2}+\mathfrak{\tilde{m}}_{z}^{2}}}{\left(\lambda^{2}+\mathfrak{\tilde{m}}_{z}^{2}\right)^{3/2}}$\tabularnewline
\hline 
$\,$$\,$$\Delta\neq0$$\,$$\,$ & $\mathfrak{\tilde{m}}_{z}+\Delta$ & $\sqrt{\left(\Delta-\mathfrak{\tilde{m}}_{z}\right)^{2}+4\lambda^{2}}$ & $\frac{\lambda(\Delta+\tilde{\mathfrak{m}}_{z})}{\sqrt{\lambda^{2}+\mathfrak{\tilde{m}}_{z}^{2}}}+\frac{\lambda\mathfrak{m}_{x}\sqrt{\mathfrak{\tilde{m}}_{z}(\Delta+\mathfrak{\mathfrak{\tilde{m}}}_{z})\left(2\lambda^{2}+\mathfrak{\tilde{m}}_{z}^{2}-\Delta\mathfrak{\mathfrak{\tilde{m}}}_{z}\right)}}{\left(\lambda^{2}+\mathfrak{\mathfrak{\tilde{m}}}_{z}^{2}\right)^{3/2}}$ & $\frac{\lambda(\Delta+\tilde{\mathfrak{m}}_{z})}{\sqrt{\lambda^{2}+\mathfrak{m}_{z}^{2}}}-\frac{\lambda\mathfrak{m}_{x}\sqrt{\tilde{\mathfrak{m}}_{z}(\Delta+\tilde{\mathfrak{m}}_{z})\left(2\lambda^{2}+\tilde{\mathfrak{m}}_{z}^{2}-\Delta\mathfrak{\tilde{m}}_{z}\right)}}{\left(\lambda^{2}+\mathfrak{\tilde{m}}_{z}^{2}\right)^{3/2}}$\tabularnewline
\hline 
\end{tabular}

\caption{\label{tab:limits}Spectral regimes of the $C_{3v}$ model. To ease
the notation, all couplings are taken to be positive.}
\end{table*}

\section*{Appendix B: Symmetry Analysis of Fermi-surface $K$-response tensor }

The Fermi surface density-current response functions are determined
from the zero temperature limit of the term I in the Kubo-Streda formula,
namely
\begin{equation}
K_{ai}=\frac{1}{2\pi}\int(d\mathbf{p})\textrm{Tr}\left\{ s_{a}\mathcal{G}_{\mathbf{p}}^{+}\tilde{J}_{i}\mathcal{G}_{\mathbf{p}}^{-}\right\} \,,\label{eq:B1}
\end{equation}
where $\tilde{J}_{i}$ is the renormalized current density vertex
and $(d\mathbf{p})\equiv\text{d}\mathbf{p}/\left(2\pi\right)^{2}$.
To determine the parity of $K_{ai}$ with respect to the field reversal
$\mathfrak{m}\rightarrow-\mathfrak{m}$, it suffices to consider the
``empty'' bubble \citep{Titov_2DEGRashba_SOT}. The disorder-averaged
Green's functions satisfy the following symmetry relations
\begin{align}
s_{x}\sigma_{y}\mathcal{G}^{a}(-p_{x},p_{y})\sigma_{y}s_{x} & =\left.\mathcal{G}^{a}(p_{x},p_{y})\right|_{S_{1}}\,,\label{eq:6}\\
s_{y}\sigma_{x}\mathcal{G}^{a}(p_{x},-p_{y})s_{y}\sigma_{x} & =\left.\mathcal{G}^{a}(p_{x},p_{y})\right|_{S_{2}}\,,\\
s_{z}\sigma_{z}\mathcal{G}^{a}(-p_{x},-p_{y})s_{z}\sigma_{z} & =\left.\mathcal{G}^{a}(p_{x},p_{y})\right|_{S_{3}}\,,\label{eq:7}
\end{align}
with $S_{1}\equiv\{\mathfrak{m}_{z}\rightarrow-\mathfrak{m}_{z},\lambda_{\textrm{sv}}\rightarrow-\lambda_{\textrm{sv}},\Delta\rightarrow-\Delta\}$,
$S_{2}\equiv\{\mathfrak{m}_{z}\rightarrow-\mathfrak{m}_{z},\mathfrak{m}_{x}\rightarrow-\mathfrak{m}_{x},\lambda_{\textrm{sv}}\rightarrow-\lambda_{\textrm{sv}},\Delta\rightarrow-\Delta\}$
and $S_{3}\equiv\{\mathfrak{m}_{x}\rightarrow-\mathfrak{m}_{x}\}$.
Using these symmetries, we find after some straightforward algebra
(sum over repeated indices $a=x,z$ is implied)\begin{widetext}

\begin{equation}
\{K_{ai}\}=\left(\begin{array}{cc}
\mathfrak{m}_{z}\kappa_{xx}+\mathfrak{m}_{a}^{2}\vec{\alpha}_{\xi}\cdot\vec{f_{xx}^{a}}+\vec{\alpha}_{\xi}\cdot\vec{g}\qquad & \kappa_{xy}+\mathfrak{m}_{z}\vec{\alpha}_{\xi}\cdot\vec{f}_{xy}+\mathfrak{m}_{a}^{2}h_{xy}^{a}\\
-\kappa_{xy}-\mathfrak{m}_{z}\vec{\alpha}_{\xi}\cdot\vec{f}_{xy}+\mathfrak{m}_{a}^{2}h_{yx}^{a} & \mathfrak{m}_{z}\kappa_{xx}+\vec{\alpha}_{\xi}\cdot\vec{g}+\mathfrak{m}_{a}^{2}\vec{\alpha}_{\xi}\cdot\vec{f_{yy}^{a}}\\
\mathfrak{m}_{x}\kappa_{zx}+\mathfrak{m}_{z}\mathfrak{m}_{x}\vec{\alpha}_{\xi}\cdot\vec{z}_{zx}\qquad & \mathfrak{m}_{z}\mathfrak{m}_{x}\kappa_{zy}+\mathfrak{m}_{x}\vec{\alpha}_{\xi}\cdot\vec{z}_{zy}
\end{array}\right),
\end{equation}
\end{widetext}where $\vec{\alpha}_{\xi}=\xi(\lambda_{\textrm{sv}},\Delta)$
and $\{\kappa_{ia},\mathbf{f}_{ij},\mathbf{f}_{ij}^{a},\mathbf{g},h_{ij}^{a},\mathbf{z}_{zi}\}_{i,j=x,y}$
are even functions of $\mathfrak{m}_{x}$ and $\mathfrak{m}_{z}$.
The terms linear in $\vec{\alpha}_{\xi}$ are activated by the breaking
of sublattice symmetry, vanishing upon the summation over the two
valleys. 

\section*{Appendix C: analytical results for weak scattering regime}

\begin{table*}
\begin{centering}
\begin{tabular}{|c|c|c|}
\hline 
 & $m_{x}=0$ & $\mathcal{O}\left(m_{x}\right)$\tabularnewline
\hline 
\hline 
$\,\,\,\tilde{J}_{x}\,\,\,$ & $\,\,\Sigma_{x}s_{0},\,\Sigma_{0}s_{y},\,\Sigma_{x}s_{z},\,\Sigma_{z}s_{y}\,\,$ & $\,\,\Sigma_{x}s_{x},\,\Sigma_{y}s_{y}\,\,$\tabularnewline
\hline 
$\tilde{J}_{y}$ & $\,\,\Sigma_{y}s_{0},\,\Sigma_{0}s_{x},\,\Sigma_{y}s_{z},\,\Sigma_{z}s_{x}\,\,$ & $\,\,\Sigma_{0}s_{0},\,\Sigma_{z}s_{0},\,\Sigma_{0}s_{z},\,\Sigma_{x}s_{y},\,\Sigma_{y}s_{x},\,\Sigma_{z}s_{z}\,\,$\tabularnewline
\hline 
\end{tabular}
\par\end{centering}
\caption{\label{tab:Matrixstructures}Gaussian matrix structures of the renormalized
current vertex. In-plane magnetic coupling ($m_{x}\protect\neq0$)
generates additional orbital-spin mixings, which are fundamental for
the accurate description of SOTs.}
\end{table*}
For 2D materials with $C_{6v}$ symmetry ($\Delta=\lambda_{\textrm{sv}}=0$),
the self energy at high carrier density reads as 
\begin{equation}
\Sigma^{\pm}\left(\epsilon\right)\simeq\mp\imath\eta\left(\boldsymbol{1}+\frac{\mathfrak{m}_{x}}{\epsilon}s_{x}+\frac{\mathfrak{m}_{z}}{\epsilon}s_{z}\right),
\end{equation}
where $\eta=nu_{0}^{2}\epsilon/(4v^{2})$ is the disorder-induced
quasiparticle broadening. There is already a significant difference
at this stage between magnetized Dirac fermions and 2DEGs since, in
the latter, the self-energy is a scalar. For 2D Dirac fermions, it
is fundamental to keep the full matrix structure of $\Sigma^{\pm}$
in order to obtain physically sensible results that comply with exact
symmetry relations of the four-point vertex function, known as Ward
identities \citep{manuel_covariant_laws}. In the main text, we presented
full nonperturbative results obtained with a numerical inversion of
the Bethe-Salpeter equation (Fig.\,\ref{fig:03}). Analytical expressions
for the \emph{weak scattering regime} can be obtained by evaluating
special subsets of diagrams \citep{MilletariFerreira16a}. Three responses,
$K_{xy}$, $K_{yx}$ and $K_{zy}$, are activated at the ``Gaussian
level'' with the standard correlator $\langle V(\mathbf{x})V(\mathbf{x}^{\prime})\rangle=nu_{0}^{2}\delta(\mathbf{x}-\mathbf{x}^{\prime})$
. To capture the antidamping responses $K_{xx},K_{yy}$ and $K_{zx}$,
one supplements the ladder series with the ``Y'' diagrams generated
by the high-order correlator $\langle V(\mathbf{x})V(\mathbf{x}^{\prime})V(\mathbf{x}^{\prime\prime})\rangle=nu_{0}^{3}\delta(\mathbf{x}-\mathbf{x}^{\prime})\delta(\mathbf{x}^{\prime}-\mathbf{x}^{\prime\prime})$.
We find for the Gaussian Fermi-surface responses:
\begin{equation}
K_{xy}=-K_{yx}=\frac{2}{v\eta}\frac{\lambda^{3}\epsilon^{2}\left(\epsilon^{2}+\mathfrak{m}_{z}^{2}\right)}{\epsilon^{4}\left(\lambda^{2}+\mathfrak{m}_{z}^{2}\right)-\mathfrak{m}_{z}^{4}\left(\epsilon^{2}-3\lambda^{2}\right)},
\end{equation}
\begin{equation}
K_{zy}=-\frac{2}{v\eta}\frac{\lambda\mathfrak{m}_{x}\mathfrak{m}_{z}}{\lambda^{2}+\mathfrak{m}_{z}^{2}}+\mathcal{O}\left(\epsilon^{-3}\right)\,,
\end{equation}
\begin{equation}
\sigma_{xx}=\sigma_{yy}=\frac{\epsilon}{\eta}\left(1-\frac{4\lambda^{2}\mathfrak{m}_{z}^{2}\left(\epsilon^{2}-2\lambda^{2}\right)}{\epsilon^{4}\left(\lambda^{2}+\mathfrak{m}_{z}^{2}\right)-\mathfrak{m}_{z}^{4}\left(\epsilon^{2}-3\lambda^{2}\right)}\right)\,,
\end{equation}
and
\begin{equation}
K_{xx}^{\text{Y}}=K_{yy}^{\text{Y}}=\frac{u_{0}\mathfrak{m}_{z}\lambda^{5}\epsilon^{2}\left(\epsilon^{2}-\mathfrak{m}_{z}^{2}\right)\left(\epsilon^{2}+\mathfrak{m}_{z}^{2}\right)^{2}}{v^{3}\eta\left(\epsilon^{4}\left(\mathfrak{m}_{z}^{2}+\lambda^{2}\right)-\mathfrak{m}_{z}^{4}\left(\epsilon^{2}-3\lambda^{2}\right)\right)^{2}},\label{eq:coll}
\end{equation}
\begin{equation}
K_{zx}^{\text{Y}}=-\frac{u_{0}\lambda^{5}\mathfrak{m}_{x}\mathfrak{m}_{z}^{2}}{2v^{3}\eta\left(\lambda^{2}+\mathfrak{m}_{z}^{2}\right)^{3}}+\mathcal{O}\left(\epsilon^{-3}\right),
\end{equation}
\begin{alignat}{1}
\sigma_{xy}^{\text{Y}} & =-\sigma_{yx}^{\text{Y}}=-\frac{2u_{0}\mathfrak{m}_{z}\lambda^{6}\epsilon\left(\epsilon^{2}+\mathfrak{m}_{z}^{2}\right)^{3}}{v^{2}\eta\left(\epsilon^{4}\left(\lambda^{2}+\mathfrak{m}_{z}^{2}\right)-\mathfrak{m}_{z}^{4}\left(\epsilon^{2}-3\lambda^{2}\right)\right)^{2}}\,,
\end{alignat}
for the skew-scattering Fermi-surface terms.

The linear response function at the Gaussian level is determined by
a \emph{single} component of the renormalized vertex \citep{manuel_MDPI},
which provides a transparent scheme to identify candidate nonzero
responses $K_{ia}$ based on a symmetry analysis (see table \ref{tab:Matrixstructures}).
The current vertex transverse to the in-plane magnetic coupling axis
($\tilde{J}_{y}$) displays a complex structure with a term proportional
to $s_{z}$, showing that $K_{zy}$ is finite already at the Gaussian
level, in agreement with the analysis of Boltzmann transport equations
outlined in the main text.

\begin{figure}[H]
\begin{centering}
\includegraphics[scale=0.4]{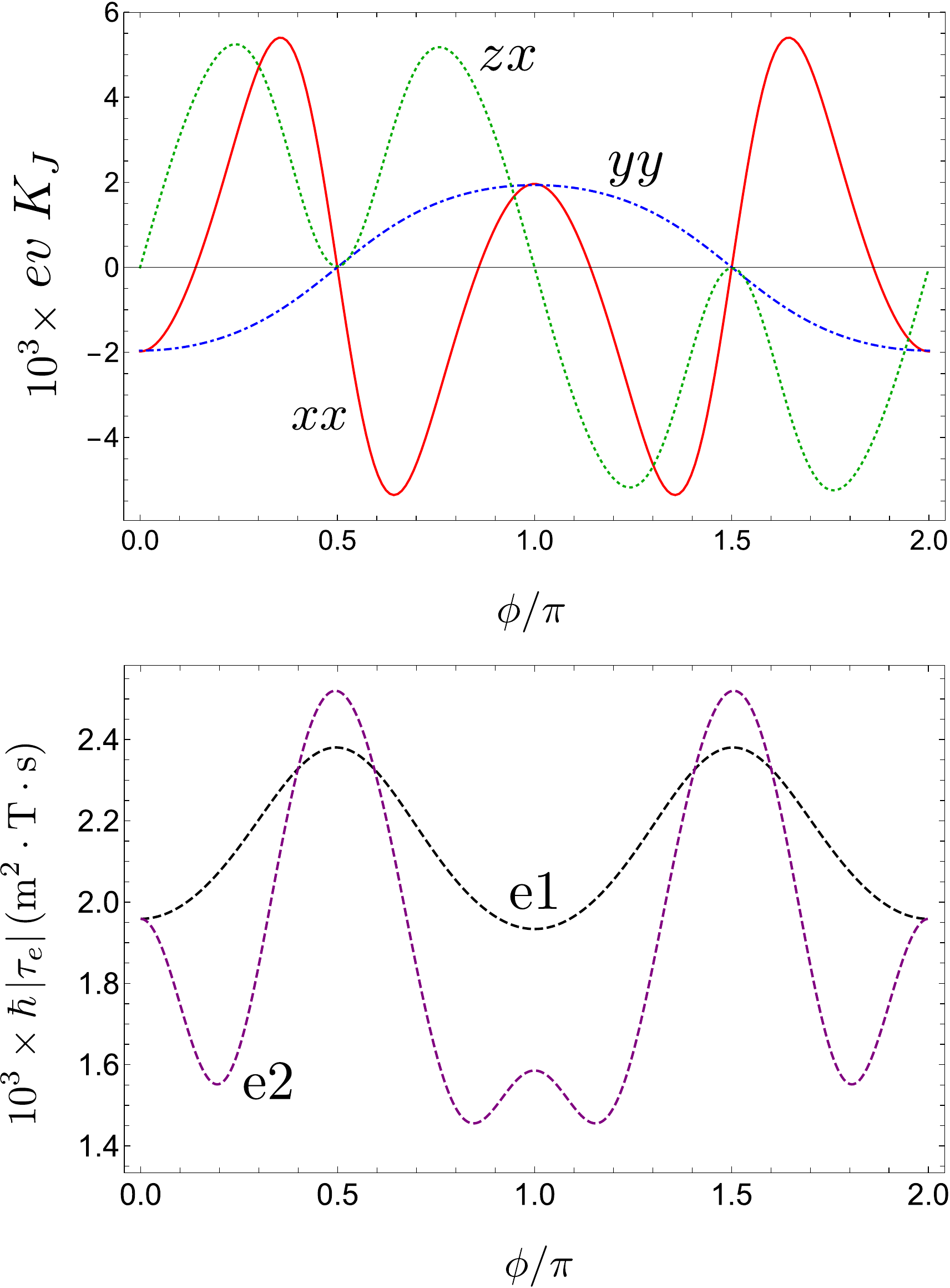}
\par\end{centering}
\caption{\label{fig:fig_4}\textit{ }Angular dependence of $\mathbf{m}$-odd
response functions (top) and associated SOT efficiencies (bottom).
System parameters: $\Delta_{\textrm{xc}}=0.1\,\text{eV}$, $\lambda=60\text{meV},$
$\lambda_{\textrm{sv}}=\Delta=0$, $n=10^{11}\text{cm}^{-2}$ and
$\epsilon=0.4$ eV.}
\end{figure}

Defining $\gamma_{ij}\equiv\Sigma_{i}s_{j}$, one obtains $\tilde{\gamma}_{ij}=\sum_{kl}c_{ijkl}\gamma_{kl}$
with the following nonzero coefficients at the Gaussian level
\begin{equation}
c_{1010}=c_{2020}=2-\frac{4\lambda^{2}\mathfrak{m}_{z}^{2}\left(\epsilon^{2}-2\lambda^{2}\right)}{\epsilon^{4}\left(\lambda^{2}+\mathfrak{m}_{z}^{2}\right)-\mathfrak{m}_{z}^{4}\left(\epsilon^{2}-3\lambda^{2}\right)},\label{eq:Long_cond}
\end{equation}
\begin{equation}
c_{1002}=-c_{2001}=-\frac{2\lambda^{3}\epsilon\left(\epsilon^{2}+\mathfrak{m}_{z}^{2}\right)}{\epsilon^{4}\left(\lambda^{2}+\mathfrak{m}_{z}^{2}\right)-\mathfrak{m}_{z}^{4}\left(\epsilon^{2}-3\lambda^{2}\right)},\label{eq:edelstein}
\end{equation}
\begin{equation}
c_{2003}=-\frac{2\lambda\mathfrak{m}_{x}\mathfrak{m}_{z}}{\epsilon\left(\lambda^{2}+\mathfrak{m}_{z}^{2}\right)}+\mathcal{O}\left(\epsilon^{-2}\right).\label{eq:zy}
\end{equation}
The longitudinal conductivity is determined by Eq. (\ref{eq:Long_cond}),
the Edelstein (inverse spin galvanic) effect is encoded in Eq. (\ref{eq:edelstein}),
and Eq. (\ref{eq:zy}) determines the generation of out-of-plane nonequilibrium
spin polarization. Replacing the renormalized vertex in the \textquotedblleft bubble\textquotedblright{}
{[}Eq. (\ref{eq:B1}){]} yields the Gaussian response functions presented
earlier. 

\section*{Appendix D: Full angular dependence of current-induced SOT}

From the knowledge of the spin-charge response tensor $\hat{K}^{J}(\phi)=\hat{K}(\phi)\cdot\hat{\sigma}(\phi)^{-1}$,
one can easily extract the SOT efficiencies or ``torkances'' \begin{widetext}
\begin{align}
t_{\textrm{o}1}(\phi)= & K_{xy}^{J}(\phi)-\tan(\phi)\,K_{zy}^{J}(\phi)\,,\label{eq:t_odd_12-1}\\
t_{\textrm{o}2}(\phi)= & \textrm{cosec}^{2}(\phi)\left[K_{xy}^{J}(\phi)+K_{yx}^{J}(\phi)\right]-2\textrm{cosec}(2\phi)K_{zy}^{J}(\phi)\,,\\
t_{\textrm{e}1}(\phi)= & \sec(\phi)K_{yy}^{J}(\phi)\,,\quad\quad\quad\label{eq:t_even_12-1}\\
t_{\textrm{e}2}(\phi)= & \textrm{cosec}^{2}(\phi)\sec(\phi)\left[K_{xx}^{J}(\phi)-K_{yy}^{J}(\phi)\right]-\textrm{cosec}(\phi)K_{zx}^{J}(\phi)-\sec(\phi)K_{xx}^{J}(\phi)\,,
\end{align}
\end{widetext} entering in the final expression for the current-induced
SOT\begin{widetext}
\begin{equation}
\mathbf{T}=t_{\textrm{o}1}(\phi)\,\mathbf{m}\times(\hat{z}\times\mathbf{J})+t_{\textrm{e}1}(\phi)\,\mathbf{m}\times(\mathbf{m}\times(\hat{z}\times\mathbf{J}))+t_{\textrm{o}2}(\phi)\,\mathbf{m}\times(\mathbf{m}\times\hat{z})(\mathbf{m}\cdot\mathbf{J})+t_{\textrm{e}2}(\phi)\,\mathbf{m}\times\hat{z}\,(\mathbf{m}\cdot\mathbf{J})\,.\label{eq:2}
\end{equation}
\end{widetext}

These expressions match those reported by I. A. Ado \emph{et al} \citep{Titov_2DEGRashba_SOT}
(apart from an overall minus sign in the $\mathbf{m}-$even torques).
The evaluation fo the angular dependence of the torkance funcions
$\{t_{\textrm{o}1}(\phi),t_{\textrm{o}2}(\phi),t_{\textrm{e}1}(\phi),t_{\textrm{e}2}(\phi)\}$
requires a full nonperturbative treatment beyond previous microscopic
formulations\textcolor{black}{{} \citep{Barnas_SOT_Graphene_PRB2015,Titov_2DEGRashba_SOT,TI_SOT,PhysRevLett.108.117201}}.
The angular dependence of the skew scattering-activated response function
and SOT efficiency parameters obtained by a numerically exact procedure
is depicted in Fig. \ref{fig:fig_4} for a magnetized graphene layer
with strong interfacial exchange and SOC effects. 

\bibliographystyle{unsrt}
\bibliography{ref}

\begin{thebibliography}{10}

\bibitem{RevModPhys.91.035004}
A.~Manchon and et~al.
\newblock Current-induced spin-orbit torques in ferromagnetic and
  antiferromagnetic systems.
\newblock {\em Rev. Mod. Phys.}, 91:035004, 2019.

\bibitem{Nature_SOT_magnetic_switching}
I.~M. Miron and et~al.
\newblock Perpendicular switching of a single ferromagnetic layer induced by
  in-plane current injection.
\newblock {\em Nature}, 476, 2011.

\bibitem{Nature_SOT_SHE_magnetic_switching}
C.~O. Avci and et~al.
\newblock Current-induced switching in a magnetic insulator.
\newblock {\em Nature Materials}, 16:309, 2017.

\bibitem{Fast_Magnetization_Switching}
S.~Shi, Y.~Ou, S.~V. Aradhya, D.~C. Ralph, and R.~A. Buhrman.
\newblock Fast low-current spin-orbit-torque switching of magnetic tunnel
  junctions through atomic modifications of the free-layer interfaces.
\newblock {\em Phys. Rev. Applied}, 9:011002, 2018.

\bibitem{LLG_equation}
T.~L. Gilbert.
\newblock A phenomenological theory of damping in ferromagnetic materials.
\newblock {\em IEEE Transactions on Magnetics}, 40:3, 2004.

\bibitem{Garello13}
K.~Garello and et~al.
\newblock Symmetry and magnitude of spin--orbit torques in ferromagnetic
  heterostructures.
\newblock {\em Nature Nanotechnology}, 8(8):587, 2013.

\bibitem{Titov_2DEGRashba_SOT}
I.~A. Ado, Oleg~A. Tretiakov, and M.~Titov.
\newblock Microscopic theory of spin-orbit torques in two dimensions.
\newblock {\em Phys. Rev. B}, 95:094401, 2017.

\bibitem{2D_CrI3_Magnetic_Layer_Dependent}
B.~Huang and et~al.
\newblock Layer-dependent ferromagnetism in a van der waals crystal down to the
  monolayer limit.
\newblock {\em Nature}, 546:270, 2017.

\bibitem{2D_Ferromagnetism_CrGeTe_vdW}
C.~Gong and et~al.
\newblock Discovery of intrinsic ferromagnetism in two-dimensional van der
  waals crystals.
\newblock {\em Nature}, 546:265, 2017.

\bibitem{SOT_Switching_Fe3GeTe2}
M.~Alghamdi and et~al.
\newblock Highly efficient spin-orbit torque and switching of layered
  ferromagnet {F}e$_3${G}e{T}e$_2$.
\newblock {\em Nano Letters}, 19(7):4400, 2019.

\bibitem{SOT_Switching_Fe3GeTe2_SciAdv}
Xiao Wang and et~al.
\newblock Current-driven magnetization switching in a van der waals ferromagnet
  {F}e$_3${G}e{T}e$_2$.
\newblock {\em Science Advances}, 5, 2019.

\bibitem{SOT2D_WTe2_17}
MacNeill D. and et~al.
\newblock Control of spin-orbit torques through crystal symmetry in wete2
  ferromagnet bilayers.
\newblock {\em Nature Phys.}, 13:300, 2017.

\bibitem{SOT2D_WTe2_17_b}
D.~MacNeill and et~al.
\newblock Thickness dependence of spin-orbit torques generated by
  ${\text{wte}}_{2}$.
\newblock {\em Phys. Rev. B}, 96:054450, 2017.

\bibitem{SOT2D_WeTe2_19}
S.~Shi and et~al.
\newblock All-electric magnetization switching and dzyaloshinskii-moriya
  interaction in wte2-ferromagnet heterostructures.
\newblock {\em Nat. Nanotech.}, 14:945, 2019.

\bibitem{SOT_2DvdW_device_abinitio}
Kapildeb Dolui and et~al.
\newblock Proximity spin-orbit torque on a two-dimensional magnet within van
  der waals heterostructure: Current-driven antiferromagnet-to-ferromagnet
  reversible nonequilibrium phase transition in bilayer cri3.
\newblock {\em Nano Letters}, 20(4):2288--2295, 2020.

\bibitem{SOT_2DvdW_graphene_device_abinitio}
K.~Zollner and et~al.
\newblock Purely interfacial and highly tunable by gate or disorder spin-orbit
  torque in graphene doubly proximitized by two-dimensional ferromagnet
  cr2ge2te6 and monolayer ws2.
\newblock {\em arXiv:1910.08072 [cond-mat.mes-hall]}.

\bibitem{PhysRevB.94.054415}
S.~Wimmer, K.~Chadova, M.~Seemann, D.~K\"odderitzsch, and H.~Ebert.
\newblock Fully relativistic description of spin-orbit torques by means of
  linear response theory.
\newblock {\em Phys. Rev. B}, 94:054415, Aug 2016.

\bibitem{Hernando06}
D.~Huertas-Hernand, F.~Guinea, and A.~Brataas.
\newblock Spin-orbit coupling in curved graphene, fullerenes, nanotubes, and
  nanotube caps.
\newblock {\em Phys. Rev. B}, 74:155426, 2006.

\bibitem{Kochan17}
D.~Kochan, S.~Irmer, and J.~Fabian.
\newblock Model spin-orbit coupling hamiltonians for graphene systems.
\newblock {\em Phys. Rev. B}, 95:165415, 2017.

\bibitem{MilletariFerreira16a}
M.~Milletar\`{\i} and A.~Ferreira.
\newblock Quantum diagrammatic theory of the extrinsic spin {H}all effect in
  graphene.
\newblock {\em Phys. Rev. B}, 94:134202, 2016.

\bibitem{MilletariFerreira16b}
M.~Milletar\`{\i} and A.~Ferreira.
\newblock Crossover to the anomalous quantum regime in the extrinsic spin hall
  effect of graphene.
\newblock {\em Phys. Rev. B}, 94:201402, 2016.

\bibitem{manuel_covariant_laws}
M.~Milletar\`{\i}, M.~Offidani, A.~Ferreira, and R.~Raimondi.
\newblock Covariant conservation laws and the spin hall effect in dirac-rashba
  systems.
\newblock {\em Phys. Rev. Lett.}, 119:246801, 2017.

\bibitem{manuel_optimalS2C}
M.~Offidani, M.~Milletar\`{\i}, R.~Raimondi, and A.~Ferreira.
\newblock Optimal charge-to-spin conversion in graphene on transition-metal
  dichalcogenides.
\newblock {\em Phys. Rev. Lett.}, 119:196801, Nov 2017.

\bibitem{Tokatly08}
I.~V. Tokatly.
\newblock Equilibrium spin currents: Non-abelian gauge invariance and color
  diamagnetism in condensed matter.
\newblock {\em Phys. Rev. Lett.}, 101:106601, 2008.

\bibitem{TMD1}
Z.~Y. Zhu, Y.~C. Cheng, and U.~Schwingenschl\"ogl.
\newblock Giant spin-orbit-induced spin splitting in two-dimensional
  transition-metal dichalcogenide semiconductors.
\newblock {\em Phys. Rev. B}, 84:153402, 2011.

\bibitem{TMD2}
Di~Xiao, Gui-Bin Liu, Wanxiang Feng, Xiaodong Xu, and Wang Yao.
\newblock Coupled spin and valley physics in monolayers of ${\mathrm{mos}}_{2}$
  and other group-vi dichalcogenides.
\newblock {\em Phys. Rev. Lett.}, 108:196802, 2012.

\bibitem{Rashba_Bychkov_SOC}
A.~Bychkov and E.~I. Rashba.
\newblock Properties of a 2{D} electron gas with lifted spectral degeneracy.
\newblock {\em JETP Lett.}, 39, 1984.

\bibitem{Rashba09}
Emmanuel~I. Rashba.
\newblock Graphene with structure-induced spin-orbit coupling: Spin-polarized
  states, spin zero modes, and quantum hall effect.
\newblock {\em Phys. Rev. B}, 79:161409, 2009.

\bibitem{Kane_Mele_QSHE}
C.~L. Kane and E.~J. Mele.
\newblock Quantum spin {H}all effect in graphene.
\newblock {\em Phys. Rev. Lett.}, 95, Nov 2005.

\bibitem{Cummings17}
A.~W. Cummings, J.~H. Garcia, J.~Fabian, and S.~Roche.
\newblock Giant spin lifetime anisotropy in graphene induced by proximity
  effects.
\newblock {\em Phys. Rev. Lett.}, 119:206601, 2017.

\bibitem{Manuel_SpinRelax_GrapheneTMD}
M.~Offidani and A.~Ferreira.
\newblock Microscopic theory of spin relaxation anisotropy in graphene with
  proximity-induced spin-orbit coupling.
\newblock {\em Phys. Rev. B}, 98:245408, 2018.

\bibitem{Ghiasi17}
Talieh~S. Ghiasi, Josep Ingla-Aynes, Alexey~A. Kaverzin, and Bart~J. van Wees.
\newblock Large proximity-induced spin lifetime anisotropy in transition-metal
  dichalcogenide/graphene heterostructures.
\newblock {\em Nano Letters}, 17(12):7528--7532, 2017.
\newblock PMID: 29172543.

\bibitem{Benitez18}
L.~A. Ben{\'\i}tez and et~al.
\newblock Strongly anisotropic spin relaxation in graphene--transition metal
  dichalcogenide heterostructures at room temperature.
\newblock {\em Nature Physics}, 14:303, 2018.

\bibitem{AHE_Manuel_Dirac}
M.~Offidani and A.~Ferreira.
\newblock Anomalous hall effect in 2d dirac materials.
\newblock {\em Phys. Rev. Lett.}, 121:126802, 2018.

\bibitem{Burkov04}
A.~A. Burkov, Alvaro~S. N\'u\~nez, and A.~H. MacDonald.
\newblock Theory of spin-charge-coupled transport in a two-dimensional electron
  gas with rashba spin-orbit interactions.
\newblock {\em Phys. Rev. B}, 70:155308, 2004.

\bibitem{Garate10}
I.~Garate and M.~Franz.
\newblock Inverse spin-galvanic effect in the interface between a topological
  insulator and a ferromagnet.
\newblock {\em Phys. Rev. Lett.}, 104:146802, 2010.

\bibitem{Yokoyama10}
Takehito Yokoyama, Jiadong Zang, and Naoto Nagaosa.
\newblock Theoretical study of the dynamics of magnetization on the topological
  surface.
\newblock {\em Phys. Rev. B}, 81:241410, 2010.

\bibitem{Shen14}
Ka~Shen, G.~Vignale, and R.~Raimondi.
\newblock Microscopic theory of the inverse edelstein effect.
\newblock {\em Phys. Rev. Lett.}, 112:096601, 2014.

\bibitem{TI_SOT}
P.~B. Ndiaye, C.~A. Akosa, M.~H. Fischer, A.~Vaezi, E.-A. Kim, and A.~Manchon.
\newblock Dirac spin-orbit torques and charge pumping at the surface of
  topological insulators.
\newblock {\em Phys. Rev. B}, 96:014408, 2017.

\bibitem{Note1}
To simplify the analysis, we neglect the $\theta $-dependence of the transport
  times. This is justified since the anisotropy present in the equilibrium spin
  texture suffices to capture the qualitative behavior of the density-current
  response function.

\bibitem{Note2}
This phenomenon is distinct from quantized magnetoelectric effect inside the
  surface gap of topological insulator/ferromagnet interfaces, $\protect
  \mathbf {S}=-\sigma _{H}\protect \mathbf {E}$ with $\sigma _{H}$ the Hall
  conductance, which is a topological effect \protect \citep {Garate10}.

\bibitem{RS_graphene}
Z.~H. Ni and et~al.
\newblock On resonant scatterers as a factor limiting carrier mobility in
  graphene.
\newblock {\em Nano Letters}, 10:3868, 2010.

\bibitem{Ferreira11rs}
A.~Ferreira and et~al.
\newblock Unified description of the dc conductivity of monolayer and bilayer
  graphene at finite densities based on resonant scatterers.
\newblock {\em Phys. Rev. B}, 83:165402, Apr 2011.

\bibitem{AHE_Kubo_Streda}
A.~Cr\'{e}pieux and P.~Bruno.
\newblock Theory of the anomalous hall effect from the kubo formula and the
  dirac equation.
\newblock {\em Phys. Rev. B}, 64:014416, 2001.

\bibitem{Pesin12}
D.~A. Pesin and A.~H. MacDonald.
\newblock Quantum kinetic theory of current-induced torques in rashba
  ferromagnets.
\newblock {\em Phys. Rev. B}, 86:014416, Jul 2012.

\bibitem{Federov13}
Dmitry~V. Fedorov and et~al.
\newblock Impact of electron-impurity scattering on the spin relaxation time in
  graphene: A first-principles study.
\newblock {\em Phys. Rev. Lett.}, 110:156602, Apr 2013.

\bibitem{Ferreira14}
A.~Ferreira, T.~G. Rappoport, M.~A. Cazalilla, and A.~H.~C. Castro.
\newblock Extrinsic spin hall effect induced by resonant skew scattering in
  graphene.
\newblock {\em Phys. Rev. Lett.}, 112:066601, 2014.

\bibitem{Pachoud14}
A.~Pachoud, A.~Ferreira, B.~\"Ozyilmaz, and A.~H.~C. Neto.
\newblock Scattering theory of spin-orbit active adatoms on graphene.
\newblock {\em Phys. Rev. B}, 90:035444, 2014.

\bibitem{Huang14}
C.~Huang, Y.~D. Chong, and M.~A. Cazalilla.
\newblock Direct coupling between charge current and spin polarization by
  extrinsic mechanisms in graphene.
\newblock {\em Phys. Rev. B}, 94:085414, 2016.

\bibitem{Note3}
A. Veneri, F. Sousa, and A. Ferreira (to be published).

\bibitem{manuel_MDPI}
M.~Offidani, R.~Raimondi, and A.~Ferreira.
\newblock Microscopic linear response theory of spin relaxation and
  relativistic transport phenomena in graphene.
\newblock {\em Condens. Matter}, 3:18, 2018.

\bibitem{SGE_2D_Ghiasi}
T.~S. Ghiasi, A.~A. Kaverzin, P.~J. Blah, and B.~J. van Wees.
\newblock Charge-to-spin conversion by the rashba-edelstein effect in
  two-dimensional van der waals heterostructures up to room temperature.
\newblock {\em Nano Letters}, 19:5959, 2019.

\bibitem{SGE_2D_Benitez}
L.~A. Ben{\'\i}tez and et~al.
\newblock Tunable room-temperature spin galvanic and spin hall effects in van
  der waals heterostructures.
\newblock {\em Nature Materials}, 19:170, 2020.

\bibitem{SGE_2D_Lin20}
L.~Lin and et~al.
\newblock Gate-tunable reversible rashba-edelstein effect in a few-layer
  graphene/2h-tas2 heterostructure at room tempertature.
\newblock {\em ACS Nano}, 2020 (in press).

\bibitem{K2020}
Dmitrii Khokhriakov, Anamul~Md. Hoque, Bogdan Karpiak, and Saroj~P. Dash.
\newblock Gate-tunable spin-galvanic effect in graphene-topological insulator
  van der waals heterostructures at room temperature.
\newblock {\em Nature Communications}, 11(1):3657, 2020.

\bibitem{TMD_giantMEC1}
Jingshan Qi, Xiao Li, Qian Niu, and Ji~Feng.
\newblock Giant and tunable valley degeneracy splitting in
  ${\mathrm{mote}}_{2}$.
\newblock {\em Phys. Rev. B}, 92:121403, 2015.

\bibitem{TMDgiantMEC2}
Qun-Fang Yao and et~al.
\newblock Manipulation of the large rashba spin splitting in polar
  two-dimensional transition-metal dichalcogenides.
\newblock {\em Phys. Rev. B}, 95:165401, 2017.

\bibitem{TMDgiantMEC3}
Chuan Zhao and et~al.
\newblock Enhanced valley splitting in monolayer wse2 due to magnetic exchange
  field.
\newblock {\em Nature Nanotechnology}, 12:757, 2017.

\bibitem{TMDgiantMEC4}
T.~Norden and et~al.
\newblock Giant valley splitting in monolayer ws2 by magnetic proximity effect.
\newblock {\em Nat.Commun.}, 10(1):4163, 2019.

\bibitem{PhysRevB.101.020407}
K.~D. Belashchenko, Alexey~A. Kovalev, and M.~van Schilfgaarde.
\newblock Interfacial contributions to spin-orbit torque and magnetoresistance
  in ferromagnet/heavy-metal bilayers.
\newblock {\em Phys. Rev. B}, 101:020407, Jan 2020.

\bibitem{PhysRevB.102.014401}
Fei Xue, Christoph Rohmann, Junwen Li, Vivek Amin, and Paul Haney.
\newblock Unconventional spin-orbit torque in transition metal
  dichalcogenide--ferromagnet bilayers from first-principles calculations.
\newblock {\em Phys. Rev. B}, 102:014401, Jul 2020.

\bibitem{Barnas_SOT_Graphene_PRB2015}
A.~Dyrda\l{} and J.~Barna\ifmmode~\acute{s}\else \'{s}\fi{}.
\newblock Current-induced spin polarization and spin-orbit torque in graphene.
\newblock {\em Phys. Rev. B}, 92:165404, 2015.

\bibitem{PhysRevLett.108.117201}
Xuhui Wang and Aurelien Manchon.
\newblock Diffusive spin dynamics in ferromagnetic thin films with a rashba
  interaction.
\newblock {\em Phys. Rev. Lett.}, 108:117201, Mar 2012.

\end{thebibliography}

\end{document}